\documentclass[a4paper]{article}

\author{G. Gubbiotti\footnote{e-mail:
gubbiotti@mat.uniroma3.it},$\quad$  C. Scimiterna\footnote{e-mail:
scimiterna@fis.uniroma3.it}, $\quad$ D. Levi\footnote{e-mail:
decio.levi@roma3.infn.it}
\\
 Dipartimento di Matematica e Fisica, Universita' degli Studi Roma Tre,\\ e Sezione INFN di Roma Tre,\\ Via della Vasca Navale 84, 00146 Roma (Italy)
 }\title{Algebraic entropy, symmetries and linearization of   quad equations  consistent on the cube.}
\usepackage{rotating}
\usepackage[toc]{appendix}
\usepackage{cite, amssymb}
\usepackage{braket}
\usepackage{amsmath}
\usepackage{booktabs}
\usepackage{amscd}
\usepackage[Q=yes]{examplep}
\usepackage{tikz}
\usetikzlibrary{arrows}
\usetikzlibrary{patterns}
\usetikzlibrary{calc}
\usepackage{multirow}
\usepackage{rotating}
\usepackage{array}
\usepackage{longtable}
\usepackage{pdflscape}

\renewcommand{\epsilon}{\varepsilon}
\newcommand{\Hvier}{$H^{4}$}
\newcommand{\Hsechs}{$H^{6}$}
\numberwithin{equation}{section}
\newcommand{\Z}{\mathbb{Z}}
\newcommand{\N}{\mathbb{N}}
\newcommand{\Cn}{\mathbb{C}}
\newcommand{\CP}{\mathbb{CP}}
\newcommand{\SymPy}{\texttt{SymPy}}

\newcommand{\abs}[1]{\left|#1\right|}
\DeclareMathOperator{\Mob}{M\ddot{o}b}
\DeclareMathOperator{\PGL}{PGL}

\def\bea{\begin{eqnarray}}
\def\eea{\end{eqnarray}}
\def\ri{{\rm{i}}}

\newcommand{\Fp}[1]{F^{(+)}_{#1}}
\newcommand{\Fm}[1]{F^{(-)}_{#1}}

\newcommand{\Fppp}{\Fp{n}\Fp{m}}
\newcommand{\Fpmm}{\Fp{n}\Fm{m}}
\newcommand{\Fmpm}{\Fm{n}\Fp{m}}
\newcommand{\Fmmp}{\Fm{n}\Fm{m}}

\newtheorem{theorem}{Theorem}

\allowdisplaybreaks

\begin{document}

\maketitle
\begin{abstract}
We discuss the  non autonomous nonlinear partial difference equations belonging to Boll classification of quad graph equations consistent around the cube.  We show how starting from the compatible equations on a cell we can construct the lattice equations, its B\"acklund transformations and Lax pairs. By carrying out the algebraic entropy calculations we show that the $H^4$ trapezoidal and the $H^6$ families are linearizable and in a few examples  we show how we can effectively linearize them. 
\end{abstract}
\section{Introduction}\label{introduction}

In recent years  the integrability criteria  denoted Consistency 
Around the Cube (CAC) has been a source of many results in the 
classification of nonlinear difference equations on a quad graph. Its importance 
relays in the fact that provides B\"acklund transforms \cite{BoS,BrH,N,NW} 
and as a consequence its  zero curvature representation 
or Lax pairs. As it is well known \cite{yamilov2006}  Lax 
pairs and B\"acklund transforms are associated to both linearizable and integrable equations.  While in the integrable case the solution of the Lax pair provides genuine nontrivial solutions \cite{cd1} in the linearizable case the Lax pair is fake \cite{bh1,bh2,gls16}

\begin{figure}[htbp]
   \centering
   \begin{tikzpicture}[auto]
      \node (x1) at (0,0) [circle,fill,label=-135:$x$] {};
      \node (x4) at (0,2.5) [circle,fill,label=135:$x_{1}$] {};
      \node (x2) at (2.5,0) [circle,fill,label=-45:$x_{2}$] {};
      \node (x3) at (2.5,2.5) [circle,fill,label=45:$x_{12}$] {};
      \draw [ultra thick] (x2) to node {$\alpha_{1}$} (x1);
      \draw [ultra thick] (x4) to node {$\alpha_{1}$} (x3);
      \draw [ultra thick] (x3) to node {$\alpha_{2}$} (x2);
      \draw [ultra thick] (x1) to node {$\alpha_{2}$} (x4);
   \end{tikzpicture} 
   \caption{Quad-graph}
\label{fig:bp}
\end{figure}
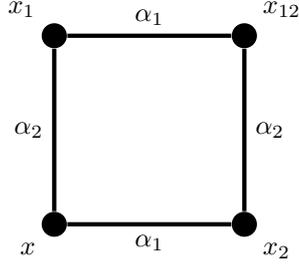

 The first attempt to carry out a classification of partial difference 
 equations using  the CAC condition  has been presented in \cite{ABS03} 
 assuming that the equations on all faces of the cube  had the same
 form. The result is a class of discrete equations   whose basic building blocks are equations on 
 quadrilaterals of the type 
 \bea 
 \label{1.1}
 A\left(x,x_{1},x_{2},x_{12};\alpha_{1},\alpha_{2}\right)=0,
 \eea
where the four fields $x$, $x_{1}$, $x_{2}$ and $x_{12}\in{\mathbb C}$ 
are assigned to the four vertexes of a quadrilateral and the parameters 
$\alpha_{i}\in{\mathbb C}$, $i=1$, $2$ to its edges, see Fig. \ref{fig:bp} (later on, in (\ref{1.4}), the points $x$, $x_1$, $x_2$ and $x_{12}$ will be denoted by $x_1$, $x_2$, $x_3$ and $x_4$).
In the notation of (\ref{1.1}), if the quadrilateral is constructed 
  from an 
origin on which the field $x$ is assigned  introducing two independent directions $1$ and $2$;  the subscript 
$1$ and $2$ denote the field in the vertex shifted by $\alpha_1$ and
$\alpha_2$ along the direction $1$ and $2$ from the origin, while $12$ 
refers to the field in the remaining vertex of the quadrilateral. 
Moreover the function $A\left(x,x_{1},x_{2},x_{12};\alpha_{1},\alpha_{2}\right)$ 
is assumed to be affine linear in each argument (multilinearity) 
with coefficients depending on the two edge parameters and invariant 
under the discrete group $D_{4}$ of the symmetries of the  square
\bea
\nonumber A\left(x,x_{1},x_{2},x_{12};\alpha_{1},\alpha_{2}\right)&=&\epsilon A\left(x,x_{2},x_{1},x_{12};\alpha_{2},\alpha_{1}\right)\\ \nonumber &=&\sigma A\left(x_{1},x,x_{12},x_{2};\alpha_{1},\alpha_{2}\right),\qquad  (\epsilon,\sigma)=\pm 1.
\eea
 \begin{figure}[htbp] 
   \centering
   \begin{tikzpicture}[auto,scale=0.8]
      \node (x) at (0,0) [circle,fill,label=-45:$x$] {};
      \node (x1) at (4,0) [circle,fill,label=-45:$x_{1}$] {};
      \node (x2) at (1.5,1.5) [circle,fill,label=-45:$x_{2}$] {};
      \node (x3) at (0,4) [circle,fill,label=-45:$x_{3}$] {};
      \node (x12) at (5.5,1.5) [circle,fill,label=-45:$x_{12}$] {};
      \node (x13) at (4,4) [circle,fill,label=-45:$x_{13}$] {};
      \node (x23) at (1.5,5.5) [circle,fill,label=-45:$x_{23}$] {};
      \node (x123) at (5.5,5.5) [circle,fill,label=-45:$x_{123}$] {};
      \node (A) at (2.75,0.75) {$A$};
      \node (Aq) at (2.75,4.75) {$\bar A$};
      \node (B) at (0.75,2.75) {$B$};
      \node (Bq) at (4.75,2.75) {$\bar B$};
      \node (C) at (2,2) {$C$};
      \node (Cq) at (3.5,3.5) {$\bar C$};
      \draw (x) to (x1) to (x12) to (x123) to (x23) to (x3) to (x);
      \draw (x3) to (x13) to (x1);
      \draw (x13) to (x123);
      \draw [dashed] (x) to (x2) to (x12);
      \draw [dashed] (x2) to (x23);
      \draw [dotted,thick] (A) to (Aq);
      \draw [dotted,thick] (B) to (Bq);
      \draw [dotted,thick] (C) to (Cq);
   \end{tikzpicture} 
   \caption{Equations on a Cube}
   \label{fig:cube}
\end{figure}
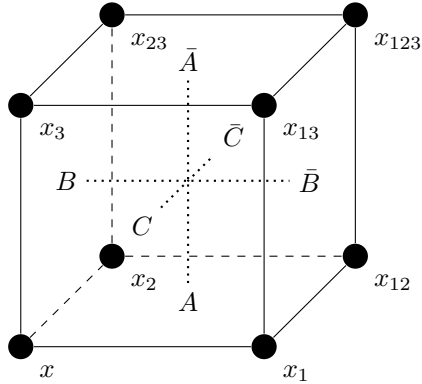
A last simplifying hypothesis is the so called \emph{tetrahedron 
property} which amounts to require that, having introduced a third independent direction $3$ and a point $x_3$ constructed in the direction $3$ at a distance $\alpha_3$ from the origin $x$,  the function 
$x_{123}\doteq x_{123}\left(x,x_{1},x_{2},x_{3};\alpha_{1},\alpha_{2},\alpha_{3}\right)$ 
 does not depend on $x$. The  result of the classification (up to a common M\"obius transformation 
of the field variables and point transformations of edge parameters) 
is given by  two lists of equations, $H$ and $Q$, for a total 
of seven consistent systems: 
\begin{subequations}
\begin{align}
&H_{1}\colon \, (x-x_{12})  (x_{2}-x_{1})  - \alpha_1  +   \alpha_2   =  0 ,\label{1.3} \\
&H_{2}\colon \, (x-x_{12})(x_{2}-x_{1}) +(\alpha_2-\alpha_1) (x+x_{2}+x_{1}+x_{1,2}) -\nonumber \\
& \qquad \qquad   - \alpha_1^2 + \alpha_2^2 = 0, \label{1.4a} \\
&H_{3}\colon  \, \alpha_1 (x x_{2}+x_{1} x_{12}) - \alpha_2 (x x_{1}+x_{2} x_{12}) + \delta (\alpha_1^2-\alpha_2^2) = 0,\label{1.5} \\
&Q_{1}\colon\,  \alpha_1 (x-x_{1}) (x_{2}- x_{12}) - \alpha_2 (x- x_{2}) (x_{1} -x_{12}) +  \nonumber \\
& \qquad \qquad + \delta^2 \alpha_1 \alpha_2 (\alpha_1-\alpha_2)= 0,\label{1.6}\\
&Q_{2}\colon\, \alpha_1 (x-x_{1}) (x_{2}- x_{12}) - \alpha_2 (x- x_{2}) (x_{1} -x_{12}) + \nonumber\\
&  \qquad \quad \quad{}+ \alpha_1 \alpha_2 (\alpha_1-\alpha_2) (x+x_{2}+x_{1}+x_{12}) -\nonumber \\
& \qquad \qquad - \alpha_1 \alpha_2 (\alpha_1-\alpha_2) (\alpha_1^2-\alpha_1 \alpha_2 + \alpha_2^2) = 0,
  \label{1.7}\\
&Q_{3}\colon\, (\alpha_2^2-\alpha_1^2) (x x_{12}+x_{2} x_{1}) + \alpha_2 (\alpha_1^2-1) (x x_{2}+\nonumber\\
&  \qquad \quad \quad{} +x_{1} x_{12})- \alpha_1 (\alpha_2^2-1) (x x_{1}+x_{12} x_{12}) -\nonumber \\
&\qquad \qquad-\frac{\delta^2 (\alpha_1^2-\alpha_2^2) (\alpha_1^2-1) (\alpha_2^2-1)}{4 \alpha_1 \alpha_2}=0 , \label{1.9}\\
&Q_{4}\colon     \,   a_0  x_{2} x_{1} x_{12}   + a_1 (x x_{2} x_{1} + x_{2} x_{1} x_{12} + x_{1} x_{12} x + \nonumber \\&    \qquad \quad \quad{}+ x_{12} x x_{2})
+a_2 (x x_{12} + x_{2} x_{1}) + \bar{a}_2 (x x_{2}+\nonumber \\&    \qquad \quad \quad{}+x_{1} x_{12})
+ \tilde{a}_2 (x x_{1}+x_{2} x_{12})
 + a_3 (x + x_{2} + x_{1} + \nonumber \\& \qquad \qquad + x_{12}) + a_4 = 0,\label{1.8}
\end{align}
\label{eqn:hq}%
\end{subequations}
where the constants $a_0$, $a_1$, $a_2$, $\bar a_2$, $\tilde a_2$, $a_3$ and $a_4$ are well defined elliptic expressions in terms of $\alpha_1$ and $\alpha_2$ \cite{ABS03}.
In (\ref{eqn:hq}) the $Q-$type equations are those where all of the six accompanying biquadratics, defined by 
\bea \label{1.4}
A^{i,j} \equiv A^{i,j}(x_i,x_j)=A_{,x_m}A_{,x_n}-A A_{,x_m x_n},
\eea
where $\{m,n\}$ is the complement of $\{i,j\}$ in $\{1,2,3,4\}$, are non degenerate, i.e. all different,   while in the $H-$type equations some of the biquadratics are degenerate.
In Fig. \ref{fig:cube} $\bar A$ represents a copy of the equation $A$ (one of the seven equations (\ref{1.3})) with $x$ substituted by $x_3$, $x_1$ by $x_{13}$, $x_2$ by $x_{23}$ and $x_{12}$ by $x_{123}$. On the faces $B$ and $\bar B$ we have a relation between a solution of $A$ and one of $\bar A$, i.e. an auto--B\"acklund transformation for $A$. By going over to projective space the auto--B\"acklund transformation will provide the Lax pair.

 In \cite{ABS09} the authors considered a more general perspective 
in the classification problem. They assumed that the faces of the 
consistency cube $A$, $B$, $C$ and $\bar A$, $\bar B$ and $\bar C$ 
could carry a priori different quad equations without assuming either
the $D_{4}$ symmetry or the tetrahedron property. 
 They considered six-tuples of (a priori different) quad equations 
 assigned to the faces of a 3D cube:
\begin{align}\label{system}
&A\left(x,x_{1},x_{2},x_{12};\alpha_1, \alpha_2\right)=0,&
&\bar{A}\left(x_{3},x_{13},x_{23},x_{123};\alpha_1, \alpha_2\right)=0,\notag\\
&B\left(x,x_{2},x_{3},x_{23};\alpha_3, \alpha_2\right)=0,&
&\bar{B}\left(x_{1},x_{12},x_{13},x_{123};\alpha_3, \alpha_2\right)=0,\\
&C\left(x,x_{1},x_{3},x_{13};\alpha_1, \alpha_3\right)=0,&
&\bar{C}\left(x_{2},x_{12},x_{23},x_{123};\alpha_1, \alpha_3\right)=0,\notag
\end{align}
see Fig. \ref{fig:cube}. Such a six-tuple is defined to be \emph{3D consistent} 
if, for arbitrary initial data $x$, $x_{1}$, $x_{2}$ and $x_{3}$,
the three values for $x_{123}$ (calculated by using $\bar{A}=0$, 
$\bar{B}=0$ and $\bar{C}=0$) coincide.
As a result in \cite{ABS09} they reobtained  the $Q-$type equations 
of \cite{ABS03} and some new  quad equations of 
type $H$ which turn out to be deformations of those present up above 
(\ref{1.3}--\ref{1.5}).
 
In  \cite{Boll11,Boll12a,Boll12b}, Boll, starting from  \cite{ABS09}, classified all the consistent  equations on the quad graph possessing the tetrahedron property without any other 
additional assumption. The results were summarized by Boll in \cite{Boll12b} in a set 
of theorems,  from Theorem {\bf 3.9} to Theorem {\bf 3.14}, 
listing all the consistent six-tuples configurations (\ref{system}) up to $(\Mob)^8$, 
the group of independent M\"obius transformations of the eight fields 
on the vertexes of the consistency three dimensional cube, see Fig \ref{fig:cube}. All these equations fall 
into three disjoint families: $Q-$type (no degenerate biquadratic), 
$H^{4}-$type (four biquadratics are degenerate) and $H^{6}-$type 
(all of the six biquadratics are degenerate).

\begin{figure}[htpb]
\centering
\begin{tikzpicture}[scale=2.5]
    \draw [pattern=north west lines,thick] (0,0) rectangle (1,1);
    \draw [pattern=north east lines,thick] (1,1) rectangle (2,2);
    \draw [pattern=horizontal lines,thick] (1,0) rectangle (2,1);
    \draw [pattern=vertical lines,thick] (0,1) rectangle (1,2);
    \foreach \x in {0,...,2}{
        \foreach \y in {0,...,2}{
            \node[draw,circle,inner sep=2pt,fill] at (\x,\y) {};
        }
    }
    \node[below left] at (0,0) {$x$};
    \node[below] at (1,0) {$x_1$};
    \node[below right] at (2,0) {$x$};
    \node[left] at (0,1) {$x_2$};
    \node[below left] at (1,1) {$x_{12}$};
    \node[right] at (2,1) {$x_2$};
    \node[above left] at (0,2) {$x$};
    \node[above] at (1,2) {$x_1$};
    \node[above right] at (2,2) {$x$};
    \node[] at (1/2,1/2) {$Q$};
    \node[] at (1/2+1,1/2) {$|Q$};
    \node[] at (1/2,1/2+1) {$\underline{Q}$};
    \node[] at (1/2+1,1/2+1) {$|\underline{Q}$};
\end{tikzpicture}
\caption{The ``four colors'' lattice}
\label{fig:elcell}
\end{figure}
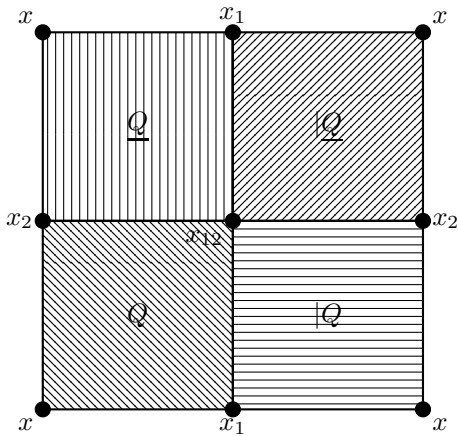

It's worth emphasizing that the classification results hold locally, 
i.e. the equations are valid on a single quadrilateral cell or on a single cube. 
The non secondary problem which has been solved is the embedding of 
the single cell/single cube equations in a $2D$/$3D$ lattice , so as to preserve the $3D$ consistency.
This was  discussed in \cite{ABS09} introducing the concept of Black--White (BW) lattice. 
To get the lattice equations one  needs to embed  (\ref{system}) into a $\Z^{2}$ lattice with an elementary cell
of dimension greater than one.  In such a case  the generic equation on a quad graph equation $Q(x,x_{1},x_{2},x_{12}; \alpha_1, \alpha_2)=0$ 
is extended to a lattice and the lattice equation will have Lax pair and B\"acklund transformation. 
To do so, following \cite{Boll11}, one reflects the square with respect 
to the normal to its right and top sides and then complete
a $2\times2$ lattice by reflecting again one of the obtained equation with respect to the
other direction\footnote{Let us note that, whatsoever side we reflect,  the result of the last reflection is the same.}. 
Such a procedure is graphically described in Fig. \ref{fig:elcell}, and at the
level of the quad equation this corresponds to constructing the three  equations obtained from
$Q=Q(x,x_{1},x_{2},x_{12};\alpha_1, \alpha_2)=0$ by flipping its fields:
\begin{subequations}
\begin{align}
Q &=Q(x,x_{1},x_{2},x_{12},\alpha_{1},\alpha_{2}) =0,
\\
|Q &=Q(x_{1},x,x_{12},x_{2},\alpha_{1},\alpha_{2}) =0,
\\
\underline{Q} &= Q(x_{2},x_{12},x,x_{1},\alpha_{1},\alpha_{2}) =0,
\\
|\underline{Q} &= Q(x_{12},x_{2},x_{1},x,\alpha_{1},\alpha_{2}) =0.
\end{align}
\label{eqn:dysys1}%
\end{subequations}
By paving the whole $\Z^{2}$ with such equations we get a partial difference equation,
which we can in principle study with the known methods. Since
\emph{a priori}  $Q\neq |Q \neq \underline{Q} \neq |\underline{Q}$  the obtained lattice will be
a four stripe lattice, i.e. an extension of the BW lattice
considered for example in \cite{HV,XP}.

Let us  notice that if the quad-equation $Q$ possess the symmetries
of the square, i.e. it is invariant under the action of $D_{4}$
one has:
\begin{equation}
    Q = |Q = \underbar{Q} = |\underbar{Q}
    \label{eq:symmsquare}
\end{equation}
implying that the elementary cell is actually of dimension one,
and one falls into the case of the ABS classification.
Beside the symmetry group of the square, $D_{4}$ \cite{ABS03} there are two others relevant
discrete symmetries for  quad equations \eqref{1.1}. 
One is the \emph{rhombic symmetry}, which holds when:
\begin{equation}
    \begin{aligned}
    Q(x,x_{1},x_{2},x_{12},\alpha_{1},\alpha_{2}) &=
    \sigma Q(x,x_{2},x_{1},x_{12},\alpha_{2},\alpha_{1})
    \\
    &=\epsilon Q(x_{12},x_{1},x_{2},x,\alpha_{2},\alpha_{1}),
    \end{aligned}
    \qquad
    (\sigma,\epsilon)\in \pm1
    \label{eq:rhombsymm}
\end{equation}
Equations with rhombic symmetries have been introduced
and classified in \cite{ABS09}.  From their explicit
form it is possible to show that they have the property:
\bea 
\label{rhombic}
Q=|\underline{Q}, \qquad \underline{Q}=|Q.
\eea
The other kind of relevant discrete symmetry for 
quad equations is the \emph{trapezoidal symmetry} \cite{ABS09} given by:
\begin{equation}
    Q(x,x_{1},x_{2},x_{12},\alpha_{1},\alpha_{2}) = Q(x_{1},x,x_{12},x_{2},\alpha_{1},\alpha_{2}).
    \label{eq:trapsymm}
\end{equation}
Eq. (\ref{eq:trapsymm}) implies:
\bea \label{trapezoidal1}
 Q=|Q, \qquad \underline{Q}=|\underline{Q}.
\eea
Geometrically the trapezoidal symmetry is an invariance
with respect to the axis parallel to ($x_{1}$, $x_{12}$). There might be a trapezoidal symmetry also with respect
to the reflection around an axe parallel to  ($x_{2}$, $x_{12}$),
but this can be reduced to the previous one by a rotation. So
there is no need to treat such symmetry but it is sufficient to consider  \eqref{trapezoidal1}.

A detailed study of all the lattices derived from the \emph{rhombic} $H^{4}$ 
family, including the construction of their three-leg forms, Lax pairs, 
B\"acklund transformations and infinite hierarchies of generalized symmetries, was presented in \cite{XP}. 

A  procedure for the  embedding of the equations defined on a cell into a 3D consistent lattice is  given in \cite{Boll11,Boll12b}. 
Different embeddings in $3D$ consistent lattices resulting either in integrable 
or non integrable equations are discussed in \cite{HV} using the algebraic entropy analysis.

 After the first results of Adler Bobenko and Suris \cite{ABS03} there have been  various 
 attempts to  reduce the requirements imposed on consistent quad equations. 
Four non tetrahedral models, three of them with $D_{4}$ symmetry, 
 were presented in \cite{H04,H05}. All of these  models turn out to be  more or less trivially linearizable \cite{RJ}. Other  consistent 
 systems of  quadrilateral lattice equations non possessing the tetrahedral property were studied in \cite{ABS09,At09}.
 
 In the following we study the independent lattice equations (consistent on the cube) 
 not already considered in the literature \cite{ABS03,ABS09,XP}, i.e. those possessing the trapezoidal symmetry or with no symmetry at all.  In Section 
 \ref{Apollo0} we list  all independent equations defined on a cell and in Section \ref{2d3dembeddings}  show in detail how 
 one  can extend the  M\"obius symmetry which classify Boll lattice equations defined on a four color lattice discussed in Appendix A. In Section \ref{Apollo1} we present all the independent 
lattice equations  obtained in this way and in Section \ref{algent} 
 we analyze them from the point of view of the algebraic entropy showing that most 
 of the new equations are linearizable. In Section \ref{Apollo2} we explicitly linearize   
  a few examples. Section \ref{concl} is devoted to some concluding remarks. 
\section{Independent equations on a single cell}\label{Apollo0}

In {\bf Theorems \boldmath{3.9}} -- {\bf 3.14}   
\cite{Boll12b},  Boll classified up to a $\left(\right.$M\"ob$\left.\right)^8$ 
symmetry every consistent six-tuples of equations with the tetrahedron property. Here we consider  all 
the  independent quad equations  
defined on a single cell not of type 
$Q$ ($Q_{1}^\epsilon$, $Q_{2}^\epsilon$, $Q_{3}^\epsilon$ and $Q_{4}$) or 
rhombic $H^{4}$ (${}_{r}H_{1}^\epsilon$, ${}_{r}H_{2}^\epsilon$ and ${}_{r}H_{3}^\epsilon$) as these two families  have been already studied extensively 
\cite{ABS03,ABS09,XP}. By independent we mean that the equations are defined  up to a $\left(\right.$M\"ob$\left.\right)^4$ 
symmetry on the fields, rotations, translations and inversions of the reference system.  
By reference system we mean those two vectors applied on the point $x$ which define the 
two oriented directions $i$ and $j$ upon which the elementary square is constructed. The vertex 
of the square lying on direction $i$ ($j$) is then indicated by $x_{i}$ ($x_{j}$). 
The remaining vertex is then called $x_{ij}$. In Fig. \ref{fig:bp} one can see an elementary square where $i=1$ and $j=2$ or viceversa.

The list we present in the following expands the analogous one given by 
{\bf Theorems \boldmath{$2.8$}-\boldmath{$2.9$}} in \cite{Boll12b}, 
where the author does not distinguish between different arrangements of the fields $x_{i}$, 
$i=1,\ldots,4$ over the four corners of the elementary square. Different 
choices reflects in different \emph{biquadratic's patterns} and, 
for any system presented in {\bf Theorems \boldmath{$2.8$}-\boldmath{$2.9$}}  
in \cite{Boll12b}, it is easy to see that a maximum of three different choices 
 may arise up to rotations, translations and inversions.      
\begin{theorem}\label{Pentalfa0}
All the independent consistent quad equations not of type 
$Q$ or \emph{rhombic} $H^{4}$ are given, up to $(\Mob)^4$ transformations 
of the fields $x$, $x_{i}$, $x_{j}$ and $x_{ij}$ and rotations, translations 
and inversions of the reference system,  by nine different representatives,
three of \Hvier-type and six of \Hsechs-type. 
We list them with their quadruples of discriminants and we identify the six-tuple
where the equation appears by the theorem number indicated in \cite{Boll12b} in the
form \textbf{3.a.b}, where \textbf{b} is the order of the six-tuple into
the theorem \textbf{3.a}.

The trapezoidal equations of type \Hvier~are:
\begin{description}
\begin{subequations}
\item[${}_{t}H_{1}^\epsilon$,  $\left(\epsilon^2,\epsilon^2,0,0\right)$:] Eq. $B$  of \textbf{ 3.10.1}.
    \begin{equation}
        \left(x-x_{2}\right)\left(x_{3}-x_{23}\right)-\alpha_{2}(1+\epsilon^2x_{3}x_{23})=0.
    \end{equation}   
\item[${}_{t}H_{2}^\epsilon$, $\left(1+4\epsilon x,1+4\epsilon x_{2},1,1\right)$:] Eq. $B$  of \textbf{ 3.10.2}.
    \begin{equation}
        \begin{aligned}
        \left(x-x_{2}\right)\left(x_{3}-x_{23}\right) &+\alpha_{2}(x+x_{2}+x_{3}+x_{23})+
        \\
        &+\frac{\epsilon\alpha_{2}}{2}\left(2x_{3}+2\alpha_{3}+\alpha_{2}\right) 
        (2x_{23}+2\alpha_{3}+\alpha_{2})+
        \\
        &+\left(\alpha_{2}+\alpha_{3}\right)^2-\alpha_{3}^2+\frac{\epsilon\alpha_{2}^3}{2}=0.
        \end{aligned}
    \end{equation}
\item[${}_{t}H_{3}^\epsilon$, $\left(x^2-4\delta^2\epsilon^2,x_{2}^2-4\delta^2\epsilon^2,x_{3}^2,x_{23}^2\right)$:] Eq. $B$  of \textbf{ 3.10.3}.
\begin{equation}
    \begin{aligned}
        e^{2\alpha_{2}}\left(xx_{23}+x_{2}x_{3}\right)&-\left(xx_{3}+x_{2}x_{23}\right)-
        \\
        &-e^{2\alpha_{3}}\left(e^{4\alpha_{2}}-1\right)
        \left(\delta^2+\frac{\epsilon^2 x_{3}x_{23}}{e^{4\alpha_{3}+2\alpha_{2}}}\right)=0.
    \end{aligned}
\end{equation}
\end{subequations}
\end{description}
The equations of type \Hsechs~are:
\begin{description}
\begin{subequations}
\item [$D_{1}$, $\left(0,0,0,0\right)$: ] Eq. $A$ of  {\bf 3.12.1} and {\bf  3.13.1}. 
\bea
 x+x_{1}+x_{2}+x_{12}=0.
\eea
This equation is invariant under any exchange of the fields.
\item[${}_{1}D_{2}$, $\left(\delta_{1}^2,\left(\delta_{1}\delta_{2}+\delta_{1}-1\right)^2,1,0\right)$:] Eq. $A$  of \textbf{ 3.12.2}.
    \begin{equation}
        \delta_{2}x+x_{1}+\left(1-\delta_{1}\right)x_{2}+x_{12}\left(x+\delta_{1}x_{2}\right)=0.
    \end{equation}
\item[$D_{3}$, $\left(4x,1,1,1\right)$:] Eq. $A$  of \textbf{ 3.12.3}.
    \begin{equation}
        x+x_{1}x_{2}+x_{1}x_{12}+x_{2}x_{12}=0.
    \end{equation}
This equation is invariant
    under the exchange $x_{1} \leftrightarrow x_{2}$.
\item[${}_{1}D_{4}$, $\left(x^2+4\delta_{1}\delta_{2}\delta_{3},x_{1}^2,x_{12}^2,x_{2}^2\right)$:]Eq. $A$  of \textbf{ 3.12.4}.
    \begin{equation}
        xx_{12}+x_{1}x_{2}+\delta_{1}x_{1}x_{12}+\delta_{2}x_{2}x_{12}+\delta_{3}=0.
    \end{equation}
  This equation is invariant
    under the simultaneous exchanges $x_{1} \leftrightarrow x_{2}$ and $\delta_{1} 
    \leftrightarrow \delta_{2}$.
\item[${}_{2}D_{2}$, $\left(\delta_{1}^2,0,1\left(\delta_{1}\delta_{2}+\delta_{1}-1\right)^2\right)$:] Eq. $C$  of \textbf{ 3.13.2}.
    \begin{equation}
        \delta_{2}x+\left(1-\delta_{1}\right)x_{3}+x_{13}+x_{1}\left(x+\delta_{1}x_{3}
        -\delta_{1}\lambda\right)-\delta_{1}\delta_{2}\lambda=0.
    \end{equation}
\item[${}_{3}D_{2}$, $\left(\delta_{1}^2,0,\left(\delta_{1}\delta_{2}+\delta_{1}-1\right)^2,1\right)$:] Eq. $C$  of \textbf{ 3.13.3}.
    \begin{equation}
        \delta_{2}x+x_{3}+\left(1-\delta_{1}\right)x_{13}
        +x_{1}\left(x+\delta_{1}x_{13}-\delta_{1}\lambda\right)-\delta_{1}\delta_{2}\lambda=0.
    \end{equation}
\item[${}_{2}D_{4}$, $\left(x^2+4\delta_{1}\delta_{2}\delta_{3},x_{1}^2,x_{2}^2,x_{12}^2\right)$:] Eq. $A$  of \textbf{ 3.13.5}.
    \begin{equation}
        xx_{1}+\delta_{2}x_{1}x_{2}+\delta_{1}x_{1}x_{12}+x_{2}x_{12}+\delta_{3}=0.
    \end{equation}
 This equation is invariant
    under the simultaneous exchanges $x_{2} \leftrightarrow x_{12}$ and $\delta_{1} 
    \leftrightarrow \delta_{2}$
\end{subequations}
\end{description}
\end{theorem}

Let us note  that differently from the rhombic
\Hvier~equations, which are $\varepsilon$-deformations of
the $H$ equations in the ABS classification \cite{ABS03} and hence, in the limit $\varepsilon \rightarrow 0$, 
have square symmetries, the trapezoidal \Hvier~equations in the limit
$\varepsilon\to0$ keep their discrete symmetry. Such class
is then completely new with respect the ABS classification and  the ``deformed'' and the ``undeformed'' equations
 share the same properties.
 
Everything is just written on a 
single cell and no dynamical system over the entire lattice exists. 
The problem of the embedding in a $2D/3D$-lattice is discussed later in Appendix A.

\section{ Transformation groups for  quad lattice equations.}
\label{2d3dembeddings}
We summarize in detail in Appendix A the results contained in \cite{Boll12b} on the construction of the  lattice equations and their Lax pairs.

Here in the following we present the extension of the M\"obius transformations necessary to treat the  equations we obtain when constructing  quad graph equations consistent on an extended four color cube.

The classification of quad equations presented in Section \ref{Apollo0}, has been carried out up to a
M\"obius transformations in each vertex:
\begin{equation}
    M\colon\left( x,x_{1},x_{2},x_{12} \right)
    \mapsto
    \left( \frac{a_0 x + b_0}{c_0 x + d_0}, \frac{a_{1} x_{1} + b_{1}}{c_{1} x_{1} + d_{1}},
    \frac{a_{2} x_{2} + b_{2}}{c_{2} x_{2} + d_{2}},\frac{a_{12} x_{12} + b_{12}}{c_{12} x_{12} + d_{12}}\right).
    \label{eqn:mob4}
\end{equation}
As in the usual M\"obius transformation we have here  
$(a_{i},b_{i},c_{i},d_{i})\in \CP^{4}\setminus V\left( a_{i}d_{i} - b_{i}c_{i} \right)\simeq \PGL(2,\Cn)$
with $i=0,1,2,12$, i.e.
 each set of parameters is defined up to
to a multiplication by a number. Obviously as the usual M\"obius transformations 
these transformations will form a group under composition 
and we shall call such group $(\Mob)^{4}$.

\begin{figure}[b!]
    \centering
    \begin{tikzpicture}[scale=3]
        \node at (0,0) {$\widehat{Q}$};
        \node at (0,1) {$Q$};
        \node at (2,1) {$M(Q)$};
        \node at (2,0) {$\widehat{M(Q)} \equiv \widehat{M}(\widehat{Q})$};
        \node at (1,-1/8) {$\widetilde{M}\in\widetilde{(\Mob)}^{4}$};
        \node at (1,1+1/8) {$M \in (\Mob)^{4}$};
        \node [below, rotate=90] at (1,1/2) {$1:1$};
        \draw [-triangle 60] plot [smooth,tension=1] coordinates {(0,-1/8) (1,-1/3) (2,-1/8)};
        \draw [-triangle 60] plot [smooth,tension=1] coordinates {(0,1+1/8) (1,1+1/3) (2,1+1/8)};
        \draw [-triangle 60] plot (0,1-1/8)--(0,1/8);
        \draw [-triangle 60] plot (2,1-1/8)--(2,1/8);
        \draw [-triangle 60] plot (1,1)--(1,0);
        \draw [-triangle 60] plot (1,0)--(1,1);
    \end{tikzpicture}
    \caption{The commutative diagram defining $\widehat{(\Mob)}^{4}$.}
    \label{fig:commutative}
\end{figure}
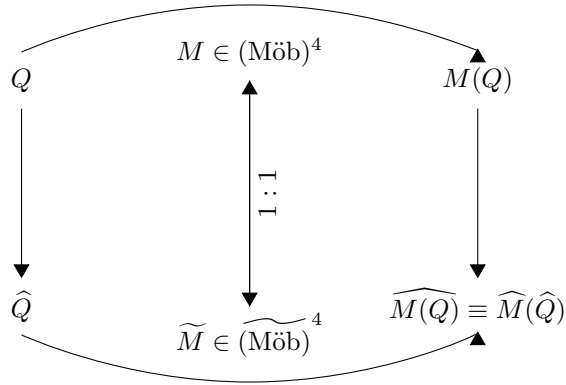

On the other hand when dealing with equations defined on the lattice we have to follow
the prescription of  Appendix \ref{nonaut}
and use the representation given by \eqref{eqn:dysys3}, i.e.
 we will have non-autonomous lattice equations.
In this Section one proves the following Theorem which extends the group $(\Mob)^{4}$ to the level of the transformations of the non autonomous lattice equations.
 We will call such group the ``non-autonomous lifting'' of $(\Mob)^{4}$,
and  denote it by $\widehat{(\Mob)}^{4}$.

 
\begin{theorem}
    \label{mobius}
    $\widehat{(\Mob)}^{4}$ is the symmetry group of the 
    non--autonomous difference equations obtained with the procedure presented in 
   Appendix \ref{nonaut} by the ``non-autonomous lifting'' of $(\Mob)^{4}$, 
    the symmetry group of the quad equations presented in Section \ref{Apollo0}.
\end{theorem}

This Theorem states that 
the result that we get by acting with a group of transformations $\widehat{(\Mob)}^{4}$ is the same of what we obtain if we first transform the equations of Section 2 using
$M\in (\Mob)^{4}$ and then we construct the non-autonomous quad equation
with the prescription of Appendix  \ref{nonaut} or viceversa if we first construct
the non-autonomous equation and then we transform it using the ``non-autonomous
lifing'' of $M$, $\widehat{M}$, see  Figure \ref{fig:commutative}.

The Theorem is made up of two parts, the proof that $\widehat{(\Mob)}^{4}$ 
is a group and that  it is equivalent to the ``non-autonomous lifting'' of $(\Mob)^{4}$.

Let us first construct, a transformation which will be the
candidate to be the ``non-autonomous lifting'' of $M\in(\Mob)^{4}$.
Given $M\in (\Mob)^{4}$ using the same ideas of Appendix \ref{nonaut}
we can construct the following transformation:
\begin{equation}
    M_{n,m} \in \widehat{(\Mob)}^{4} \colon u_{n,m} \mapsto
    \begin{aligned}[t]
        f_{n,m}\frac{a_0 u_{n,m} + b_0}{c_0 u_{n,m} + d_0} &+
        |f_{n,m}\frac{a_{1} u_{n,m} + b_{1}}{c_{1} u_{n,m} + d_{1}}+\\
        \underline{f}_{n,m}\frac{a_{2} u_{n,m} + b_{2}}{c_{2} u_{n,m} + d_{2}} &+
        |\underline{f}_{n,m}\frac{a_{12} u_{n,m} + b_{12}}{c_{12} u_{n,m} + d_{12}}.
    \end{aligned}
    \label{eqn:mobnm0}
\end{equation}
Eq. (\ref{eqn:mobnm0}) give us a mapping $\Phi$ between the group $(\Mob)^{4}$ and
a set of non-autonomous transformations of the field $u_{n,m}$. 
Moreover there is a one to one correspondence between an element $M \in (\Mob)^{4}$ (\ref{eqn:mob4}) and one $M_{n,m} \in \widehat{(\Mob)}^{4}$ (\ref{eqn:mobnm0}). $\Phi$ is given by:
\begin{subequations}
    \begin{align}
    \Phi \colon
        \begin{pmatrix}
            {\displaystyle\frac{a x + b}{c x + d}}
            \\
            {\displaystyle\frac{a_{1} x_{1} + b_{1}}{c_{1} x_{1} + d_{1}}}
        \\
        {\displaystyle\frac{a_{2} x_{2} + b_{2}}{c_{2} x_{2} + d_{2}}}
        \\
        {\displaystyle\frac{a_{12} x_{12} + b_{12}}{c_{12} x_{12} + d_{12}}}
        \end{pmatrix}^{T}
        &\longmapsto
        \begin{aligned}
            &f_{n,m}
            \frac{a u_{n,m} + b}{c u_{n,m} + d}+
            \\
            &|f_{n,m}
            \frac{a_{1} u_{n,m} + b_{1}}{c_{1} u_{n,m} + d_{1}} +
            \\
            &\underline{f}_{n,m}
            \frac{a_{2} u_{n,m} + b_{2}}{c_{2} u_{n,m} + d_{2}} +
            \\
            &|\underline{f}_{n,m}
            \frac{a_{12} u_{n,m} + b_{12}}{c_{12} u_{n,m} + d_{12}}
        \end{aligned},
        \\ \nonumber
        \mbox{ and its inverse}\\
        \Phi^{-1} \colon
        \begin{aligned}
            &f_{n,m}
            \frac{\alpha^{(0)} u_{n,m} + \beta^{(0)}}{\gamma^{(0)} u_{n,m} + \delta^{(0)}}+
            \\
            &|f_{n,m}
            \frac{\alpha^{(1)} u_{n,m} + \beta^{(1)}}{\gamma^{(1)} u_{n,m} + \delta^{(1)}} +
            \\
            &\underline{f}_{n,m}
            \frac{\alpha^{(2)} u_{n,m} + \beta^{(2)}}{\gamma^{(2)} u_{n,m} + \delta^{(2)}} +
            \\
            &|\underline{f}_{n,m}
            \frac{\alpha^{(3)} u_{n,m} + \beta^{(3)}}{\gamma^{(3)} u_{n,m} + \delta^{(3)}}
        \end{aligned}
        &\longmapsto
        \begin{pmatrix}
            {\displaystyle\frac{\alpha^{(0)} x + \beta^{(0)}}{\gamma^{(0)} x + \delta^{(0)}}}
            \\
            {\displaystyle\frac{\alpha^{(1)} x_{1} + \beta^{(1)}}{\gamma^{(1)} x_{1} + \delta^{(1)}}}
        \\
        {\displaystyle\frac{\alpha^{(2)} x_{2} + \beta^{(2)}}{\gamma^{(2)} x_{2} + \delta^{(2)}}}
        \\
        {\displaystyle\frac{\alpha^{(3)} x_{12} + \beta^{(3)}}{\gamma^{(3)} x_{12} + \delta^{(3)}}}
        \end{pmatrix}^{T}.
    \end{align}
    \label{eqn:mobmap}
\end{subequations}
We have now to prove that $\widehat{(\Mob)}^{4}$ is
a group and that the mapping  \eqref{eqn:mobmap} is
actually a group homomorphism.

  $\widehat{(\Mob)}^{4}$ 
is a subset of the general non-autonomous M\"obius transformation:
\begin{equation}
    W_{n,m}\colon u_{n,m} \mapsto \frac{a_{n,m}u_{n,m}+b_{n,m}}{c_{n,m}u_{n,m}+d_{n,m}}.
    \label{eqn:mobgen}
\end{equation}
From the general
rule of composition of two M\"obius transformations \eqref{eqn:mobgen} we get:
\begin{equation}
    W_{n,m}^{1} \left( W_{n,m}^{0} \left( u_{n,m} \right)\right)=
    \frac{(a_{n,m}^{0} a_{n,m}^{1} + b_{n,m}^{1} c_{n,m}^{0}) u_{n,m}+a_{n,m}^{1} b_{n,m}^{0} + b_{n,m}^{1} d_{n,m}^{0}}%
    {(a_{n,m}^{0} c_{n,m}^{1} + c_{n,m}^{0} d_{n,m}^{1}) u_{n,m}+b_{n,m}^{0} c_{n,m}^{1} + d_{n,m}^{0} d_{n,m}^{1}}.
    \label{eqn:mobcom}
\end{equation}
Its inverse is given by
\begin{equation}
    W_{n,m}^{-1}(u_{n,m}) = \frac{d_{n,m}u_{n,m}-b_{n,m}}{-c_{n,m}u_{n,m}+a_{n,m}},
    \label{eqn:mobgeninv}
\end{equation}

Using the computational rules given in Table \ref{tab:multf} and in (\ref{eqn:mobcom})
we find that the composition of two elements $M_{n,m}^{(1)},M_{n,m}^{(2)}\in \widehat{(\Mob)}^{4}$
with parameters $( a^{(i)}_{j},b^{(i)}_{j},c^{(i)}_{j},d^{(i)}_{j})$,
$j=0,1,2,3$ and  $i=1,2$, gives:
\begin{equation}
    M_{n,m}^{2} \left( M_{n,m}^{1} \left( u_{n,m} \right)\right)=
    \frac{
        \left\{\begin{aligned}
            \Bigl[f_{n,m} (a^{(0)}_{1} a^{(0)}_{2}+b^{(0)}_{2} c^{(0)}_{1})&+
            \\
            |f_{n,m}(a^{(1)}_{1}a^{(1)}_{2}+b^{(1)}_{2}c^{(1)}_{1})+
            \\
            \underline{f}_{n,m}(a^{(2)}_{1}a^{(2)}_{2}+b^{(2)}_{2}c^{(2)}_{1})&+
            \\
            |\underline{f}_{n,m}(a^{(3)}_{1}a^{(3)}_{2}+b^{(3)}_{2}c^{(3)}_{1})\Bigr]u_{n,m}&+
            \\
           + \Bigl[f_{n,m} (b^{(0)}_{1}a^{(0)}_{2}+d^{(0)}_{1}b^{(0)}_{2}) &+\\
                   +           |f_{n,m}(b^{(1)}_{1}a^{(1)}_{2}+d^{(1)}_{1}b^{(1)}_{2}) &+
            \\ +
            \underline{f}_{n,m}(b^{(2)}_{1}a^{(2)}_{2}+d^{(2)}_{1}b^{(2)}_{2}) &+\\+
            |\underline{f}_{n,m}(b^{(3)}_{1}a^{(3)}_{2}+d^{(3)}_{1}b^{(3)}_{2})\Bigr]
        \end{aligned}\right\}}{\left\{\begin{aligned}
            \Bigl[f_{n,m} (a^{(0)}_{1}c^{(0)}_{2}+c^{(0)}_{1}d^{(0)}_{1})&+
            \\
            |f_{n,m}(a^{(1)}_{1}c^{(1)}_{2}+c^{(1)}_{1}d^{(1)}_{1})+
            \\
            \underline{f}_{n,m}(a^{(2)}_{1}c^{(2)}_{2}+c^{(2)}_{1}d^{(2)}_{1})&+
            \\
            |\underline{f}_{n,m}(a^{(3)}_{1}c^{(3)}_{2}+c^{(3)}_{1}d^{(3)}_{1})\Bigr]u_{n,m}+
            \\
            \Bigl[f_{n,m} (b^{(0)}_{1}c^{(0)}_{2}+d^{(0)}_{1}d^{(0)}_{2}) &+ 
            \\
            |f_{n,m}(b^{(1)}_{1}c^{(1)}_{2}+d^{(1)}_{1}d^{(1)}_{2}) +
            \\
            \underline{f}_{n,m}(b^{(2)}_{1}c^{(2)}_{2}+d^{(2)}_{1}d^{(2)}_{2})&+
            \\
            |\underline{f}_{n,m}(b^{(3)}_{1}c^{(3)}_{2}+d^{(3)}_{1}d^{(3)}_{2})\Bigr]
        \end{aligned}\right\}}.
    \label{eqn:mobnmcomp}
\end{equation}
Its inverse is given by
\begin{equation}
    M_{n,m}^{-1}( u_{n,m} ) =
    \frac{
        \left\{\begin{aligned}
        &(f_{n,m} d^{(0)}+|f_{n,m} d_{1}
        +\underline{f}_{n,m} d_{2}+|\underline{f}_{n,m} d_{3})
        u_{n,m}
        -\\
        &(f_{n,m} b_{0}+|f_{n,m} b_{1}
        +\underline{f}_{n,m} b_{2}+|\underline{f}_{n,m} b_{3})
        \end{aligned}\right\}}{\left\{\begin{aligned}
        -&(f_{n,m} c_{0}+|f_{n,m}c_{1}
        +\underline{f}_{n,m} c_{2}+|\underline{f}_{n,m} c_{3})
        u_{n,m}
        +\\
        &(f_{n,m} a_{0}+|f_{n,m} a_{1}
        +\underline{f}_{n,m}a_{2}+|\underline{f}_{n,m} a_{3})
        \end{aligned}\right\}}.
    \label{eqn:mobnminv}
\end{equation}
\begin{table}[htb]
    \centering
    \begin{equation*}
        \begin{array}{ccccc}
            \toprule
            \cdot & f_{n,m} & |f_{n,m} & \underline{f}_{n,m} & |\underline{f}_{n,m}
            \\
            \midrule
            f_{n,m} & f_{n,m} & 0 & 0 & 0
            \\
            |f_{n,m} & 0 & |f_{n,m} & 0 & 0
            \\
            \underline{f}_{n,m} & 0 & 0 & \underline{f}_{n,m} & 0
            \\
            |\underline{f}_{n,m} & 0 & 0 & 0 & |\underline{f}_{n,m}
            \\
            \bottomrule
        \end{array}
    \end{equation*}
    \caption{Multiplication rules for the functions $\tilde f_{n,m}$ as given by \eqref{fsoltot}.}
    \label{tab:multf}
\end{table}
Thus one has proven that $\widehat{(\Mob)}^{4}$ is a group.

Let us   show that the maps $\Phi$ and $\Phi^{-1}$ in \eqref{eqn:mobmap}
are group homomorphism. They  preserve the identity,
and from  the formula of composition
of M\"obius transformations in $(\Mob)^{4}$ \eqref{eqn:mobnmcomp} we derive the required result.

Let us now check if  the diagram of Fig. 4 is satisfied.
Let us consider \eqref{eqn:mob4} and \eqref{eqn:mobnm0}
and a general \emph{multilinear} quad--equation:
\begin{equation}
    \begin{aligned}
    Q_{\text{gen}}\left( x,x_{1},x_{2},x_{12} \right) &= A_{0,1,2,12} x x_{1} x_{2} x_{12}+ B_{0,1,2} x x_{1} x_{2} +\\
    &+ B_{0,1,12} x x_{1} x_{12} + B_{0,2,12} x x_{2} x_{12} + B_{1,2,12} x_{1} x_{2} x_{12}\\
    &+ C_{0,1} x x_{1}+ C_{0,2} x x_{2}+ C_{0,12} x x_{12}+ C_{1,2} x_{1} x_{2}\\
    &+ C_{1,12} x_{1} x_{12} + C_{2,12} x_{2} x_{12}\\
    &+ D_{0} x + D_{1} x_{1} + D_{2} x_{2} + D_{12} x_{12} + K 
    \end{aligned}
    \label{Quadgen}
\end{equation}
where $A_{0,1,2,12}$, $B_{i,j,k}$, $C_{i,j}$, $D_{i}$ and $K$ 
with $i,j,k\in\Set{0,1,2,12}$ are arbitrary
complex constants. 
The proof that 
$\widehat{Q(M(x,x_{1},x_{2},x_{12}))} = \widehat{Q}(M_{n,m}(u_{n,m}))$
where $M \in (\Mob)^{4}$ and $M_{n,m} = \Phi (M) \in \widehat{(\Mob)}^{4}$  is a very computationally heavy calculation due to
the high number of parameters involved (twelve in the transformation\footnote{Using
 that M\"obius transformations are projectively defined  one can lower the number of parameters from sixteen to twelve,
but this implies to impose that some parameters are non-zero and thus
all various different possibilities must be taken into account.} and
fifteen in the equation \eqref{Quadgen}, twenty-seven paramrters in total) and to
the fact that rational functions are involved.
To simplify the problem it is sufficient to recall that every M\"obius
\begin{equation}
    m (z) = \frac{a z + b}{c z +d}, \quad z \in \mathbb{C}
    \label{moebius}
\end{equation}
transformation can be obtained as a superposition
of a \emph{translation}:
\begin{subequations}
\begin{equation}
    T_{a} (z) = z + a,
    \label{moebiust}
\end{equation}
\emph{dilatation}:
\begin{equation}
    D_{a} (z) = a z
    \label{moebiusd}
\end{equation}
and \emph{inversion}:
\begin{equation}
    I(z) = \frac{1}{z},
    \label{moebiusi}
\end{equation}
\end{subequations}
i.e.
\begin{equation}
    m(z) = \left(T_{a/c} \circ D_{(bc-ad)/c^{2}} \circ I \circ T_{d/c}\right)(z).
    \label{moebiusdec}
\end{equation}

As the group $(\Mob)^{4}$ is obtained by four copies
of the M\"obius group each acting on a different variable
we can decompose each entry  $M\in(\Mob)^{4}$ as
in \eqref{moebiusdec}. Therefore we need to check  $3^{4}=81$ transformations, depending at most on four parameters
. We can automatize such
proof by making a specific computer program to generate all the 
possible fundamental transformations in $(\Mob)^{4}$ and then check
them one by one reducing the computational effort. To this end
we used the Computer Algebra System (CAS) \SymPy \cite{sympy}.
This ends the proof of the Theorem.  The details of the calculations 
will be contained in \cite{gubbiotti_thesis}.

\section{Independent equations on the $2D$-lattice}\label{Apollo1}
We will now extend to the $2D$-lattice all the systems listed in 
Theorem \ref{Pentalfa0} of   Section \ref{Apollo0} according to 
the prescription given in Appendix \ref{nonaut}. Independence is now 
understood to be up to  $\widehat{(\Mob)}^{4}$ symmetry and rotations, 
translations and inversions of the reference system. The transformations 
of the reference system are take to be   acting 
 on the discrete indexes rather than on the reference frame.
For sake of compactness and as the equations are on
the lattice we shall omit the hats on the M\"obius transformations when clear.
\begin{theorem}\label{Pentalfa1} 
All the independent nonlinear, $2D$-dynamical systems not of type $Q$ 
or \emph{rhombic} $H^{4}$ which are consistent on the $3D$-lattice are given, 
up to  $\widehat{(\Mob)}^{4}$ transformations of the fields $u_{n,m}$, $u_{n,m+1}$, 
$u_{n+1,m}$ and $u_{n+1,m+1}$, rotations, translations and inversions of the 
discrete indexes $n$ and $m$, by nine non autonomous representatives,
three of trapezoidal \Hvier-type and six of \Hsechs-type.

The \Hvier~type equations are:
\begin{subequations}
    \begin{align}
        _{t}H_{1}\colon &
        \begin{aligned}[t]
        &\left(u_{n,m}-u_{n+1,m}\right) \cdot \left(u_{n,m+1}-u_{n+1,m+1}\right)-\\ 
        &-\alpha_{2}\epsilon^2\left({ F}_{m}^{\left(+\right)}u_{n,m+1}u_{n+1,m+1}
        +{ F}_{m}^{\left(-\right)}u_{n,m}u_{n+1,m}\right) -\alpha_{2}=0,
        \end{aligned}
        \label{eq:tH1e}
        \\
        _{t}H_{2}\colon &
        \begin{aligned}[t]
        &\left(u_{n,m}-u_{n+1,m}\right)\left(u_{n,m+1}-u_{n+1,m+1}\right)
        \\
        &+\alpha_{2}\left(u_{n,m}+u_{n+1,m}+u_{n,m+1}+u_{n+1,m+1}\right)
        \\
        &+\frac{\epsilon\alpha_{2}}{2} \left(2{F}_{m}^{\left(+\right)}u_{n,m+1}
        +2\alpha_{3}+\alpha_{2}\right)\left(2{F}_{m}^{\left(+\right)}u_{n+1,m+1}+2\alpha_{3}+\alpha_{2}\right)
        \\
        &+\frac{\epsilon\alpha_{2}}{2} \left(2{F}_{m}^{\left(-\right)}u_{n,m}+2\alpha_{3}
        +\alpha_{2}\right)\left(2{F}_{m}^{\left(-\right)}u_{n+1,m}+2\alpha_{3}+\alpha_{2}\right)
        \\
        &+\left(\alpha_{3}+\alpha_{2}\right)^2-\alpha_{3}^2-2\epsilon\alpha_{2}\alpha_{3}\left(\alpha_{3}+\alpha_{2}\right)=0
        \end{aligned}
        \label{eq:tH2e}
        \\
        _{t}H_{3}\colon &
        \begin{aligned}[t]
        &\alpha_{2}\left(u_{n,m}u_{n+1,m+1}+u_{n+1,m}u_{n,m+1}\right)
        \\
        &-\left(u_{n,m}u_{n,m+1}+u_{n+1,m}u_{n+1,m+1}\right)
        -\alpha_{3}\left(\alpha_{2}^{2}-1\right)\delta^2+
        \\
        &-\frac{\epsilon^2(\alpha_{2}^{2}-1)}{\alpha_{3}\alpha_{2}}
        \left({{ F}_{m}^{\left(+\right)}u_{n,m+1}u_{n+1,m+1}
        +{ F}_{m}^{\left(-\right)}u_{n,m}u_{n+1,m}}\right)=0,
        \end{aligned}
        \label{eq:tH3e}
    \end{align}
    \label{eq:trapezoidalH4}
\end{subequations}
These equations arise from the $B$ equation of the cases
\textbf{3.10.1}, \textbf{3.10.2} and \textbf{3.10.3} in \cite{Boll12b} 
respectively.

The \Hsechs~type equations are:
\begin{subequations}
    \begin{align}
    D_{1}&\colon
\begin{aligned}[t]
        &u_{n,m}+u_{n+1,m}+u_{n,m+1}+u_{n+1,m+1}=0.
\end{aligned}
        \label{eq:D1}
        \\
        _{1}D_{2} &\colon
        \begin{aligned}[t]
        &\phantom{+}\left( F_{n+m}^{\left(-\right)}-\delta_{1} F_{n}^{\left(+\right)} F_{m}^{\left(-\right)}+\delta_{2} F_{n}^{\left(+\right)} F_{m}^{\left(+\right)}\right)u_{n,m}
        \\
        &+\left( F_{n+m}^{\left(+\right)}-\delta_{1} F_{n}^{\left(-\right)} F_{m}^{\left(-\right)}+\delta_{2} F_{n}^{\left(-\right)} F_{m}^{\left(+\right)}\right)u_{n+1,m}+
        \\ 
        &+\left( F_{n+m}^{\left(+\right)}-\delta_{1} F_{n}^{\left(+\right)} F_{m}^{\left(+\right)}+\delta_{2} F_{n}^{\left(+\right)} F_{m}^{\left(-\right)}\right)u_{n,m+1}
        \\
        &+\left( F_{n+m}^{\left(-\right)}-\delta_{1} F_{n}^{\left(-\right)} F_{m}^{\left(+\right)}+\delta_{2} F_{n}^{\left(-\right)} F_{m}^{\left(-\right)}\right)u_{n+1,m+1}+
        \\ 
        &+\delta_{1}\left( F_{m}^{\left(-\right)}u_{n,m}u_{n+1,m}+ F_{m}^{\left(+\right)}u_{n,m+1}u_{n+1,m+1}\right)
        \\
        &+ F_{n+m}^{\left(+\right)}u_{n,m}u_{n+1,m+1}
        + F_{n+m}^{\left(-\right)}u_{n+1,m}u_{n,m+1}=0,
        \end{aligned}
        \label{eq:1D2}
        \\
        _{2}D_{2} &\colon
        \begin{aligned}[t]
            &\phantom{+}\left(F_{m}^{\left(-\right)}-\delta_{1}F_{n}^{\left(+\right)}F_{m}^{\left(-\right)}+\delta_{2}F_{n}^{\left(+\right)}F_{m}^{\left(+\right)}-\delta_{1} \lambda F_{n}^{\left(-\right)}F_{m}^{\left(+\right)}\right)u_{n,m}
        \\
        &+\left(F_{m}^{\left(-\right)}-\delta_{1}F_{n}^{\left(-\right)}F_{m}^{\left(-\right)}+\delta_{2}F_{n}^{\left(-\right)}F_{m}^{\left(+\right)}-\delta_{1} \lambda F_{n}^{\left(+\right)}F_{m}^{\left(+\right)}\right)u_{n+1,m}
        \\
        &+\left(F_{m}^{\left(+\right)}-\delta_{1}F_{n}^{\left(+\right)}F_{m}^{\left(+\right)}+\delta_{2}F_{n}^{\left(+\right)}F_{m}^{\left(-\right)}-\delta_{1} \lambda F_{n}^{\left(-\right)}F_{m}^{\left(-\right)}\right)u_{n,m+1}
        \\
        &+\left(F_{m}^{\left(+\right)}-\delta_{1}F_{n}^{\left(-\right)}F_{m}^{\left(+\right)}+\delta_{2}F_{n}^{\left(-\right)}F_{m}^{\left(-\right)}-\delta_{1} \lambda F_{n}^{\left(+\right)}F_{m}^{\left(-\right)}\right)u_{n+1,m+1}
        \\
        &+\delta_{1}\left(F_{n+m}^{\left(-\right)}u_{n,m}u_{n+1,m+1}+F_{n+m}^{\left(+\right)}u_{n+1,m}u_{n,m+1}\right)
        \\ 
        &+F_{m}^{\left(+\right)}u_{n,m}u_{n+1,m}+F_{m}^{\left(-\right)}u_{n,m+1}u_{n+1,m+1}
        -\delta_{1}\delta_{2}\lambda=0,
        \end{aligned}
        \label{eq:2D2}
        \\
        _{3}D_{2} &\colon
        \begin{aligned}[t]
            &\phantom{+}\left(F_{m}^{\left(-\right)}-\delta_{1}F_{n}^{\left(-\right)}F_{m}^{\left(-\right)}+\delta_{2}F_{n}^{\left(+\right)}F_{m}^{\left(+\right)}-\delta_{1} \lambda F_{n}^{\left(-\right)}F_{m}^{\left(+\right)}\right)u_{n,m}
        \\
        &+\left(F_{m}^{\left(-\right)}-\delta_{1}F_{n}^{\left(+\right)}F_{m}^{\left(-\right)}+\delta_{2}F_{n}^{\left(-\right)}F_{m}^{\left(+\right)}-\delta_{1} \lambda F_{n}^{\left(+\right)}F_{m}^{\left(+\right)}\right)u_{n+1,m}
        \\
        &+\left(F_{m}^{\left(+\right)}-\delta_{1}F_{n}^{\left(-\right)}F_{m}^{\left(+\right)}+\delta_{2}F_{n}^{\left(+\right)}F_{m}^{\left(-\right)}-\delta_{1} \lambda F_{n}^{\left(-\right)}F_{m}^{\left(-\right)}\right)u_{n,m+1}
        \\
        &+\left(F_{m}^{\left(+\right)}-\delta_{1}F_{n}^{\left(+\right)}F_{m}^{\left(+\right)}+\delta_{2}F_{n}^{\left(-\right)}F_{m}^{\left(-\right)}-\delta_{1} \lambda F_{n}^{\left(+\right)}F_{m}^{\left(-\right)}\right)u_{n+1,m+1}
        \\
        &+\delta_{1}\left(F_{n}^{\left(-\right)}u_{n,m}u_{n,m+1}+F_{n}^{\left(+\right)}u_{n+1,m}u_{n+1,m+1}\right) 
        \\ 
        &+F_{m}^{\left(-\right)}u_{n,m+1}u_{n+1,m+1}
        +F_{m}^{\left(+\right)}u_{n,m}u_{n+1,m}-\delta_{1}\delta_{2}\lambda=0,
        \end{aligned}
        \label{eq:3D2}
        \\
        D_{3} &\colon
        \begin{aligned}[t]
            &\phantom{+}F_{n}^{\left(+\right)}F_{m}^{\left(+\right)}u_{n,m}+F_{n}^{\left(-\right)}F_{m}^{\left(+\right)}u_{n+1,m}
            +F_{n}^{\left(+\right)}F_{m}^{\left(-\right)}u_{n,m+1}
            \\
            &+F_{n}^{\left(-\right)}F_{m}^{\left(-\right)}u_{n+1,m+1}
            +F_{m}^{\left(-\right)}u_{n,m}u_{n+1,m}
            \\
            &+F_{n}^{\left(-\right)}u_{n,m}u_{n,m+1}+F_{n+m}^{\left(-\right)}u_{n,m}u_{n+1,m+1}+
        \\
        &+F_{n+m}^{\left(+\right)}u_{n+1,m}u_{n,m+1}+F_{n}^{\left(+\right)}u_{n+1,m}u_{n+1,m+1}
        \\
        &+F_{m}^{\left(+\right)}u_{n,m+1}u_{n+1,m+1}=0,
        \end{aligned}
        \label{eq:D3}
        \\
        _{1}D_{4} &\colon
        \begin{aligned}[t]
            &\phantom{+}\delta_{1}\left(F_{n}^{\left(-\right)}u_{n,m}u_{n,m+1}+F_{n}^{\left(+\right)}u_{n+1,m}u_{n+1,m+1}\right)+\\
            &+\delta_{2}\left(F_{m}^{\left(-\right)}u_{n,m}u_{n+1,m}+F_{m}^{\left(+\right)}u_{n,m+1}u_{n+1,m+1}\right)+\\
            &+u_{n,m}u_{n+1,m+1}+u_{n+1,m}u_{n,m+1}+\delta_{3}=0,
        \end{aligned}
        \label{eq:1D4}
        \\
        _{2}D_{4} &\colon
        \begin{aligned}[t]
            &\phantom{+}\delta_{1}\left(F_{n}^{\left(-\right)}u_{n,m}u_{n,m+1}+F_{n}^{\left(+\right)}u_{n+1,m}u_{n+1,m+1}\right)+
            \\
            &+\delta_{2}\left(F_{n+m}^{\left(-\right)}u_{n,m}u_{n+1,m+1}+F_{n+m}^{\left(+\right)}u_{n+1,m}u_{n,m+1}\right)+
            \\
            &+u_{n,m}u_{n+1,m}+u_{n,m+1}u_{n+1,m+1}+\delta_{3}=0.
        \end{aligned}
        \label{eq:2D4}
    \end{align}
    \label{eq:h6}
\end{subequations}
The equations $_{1}D_{2}$, $_{1}D_{4}$, $_{2}D_{2}$  and $D_{3}$  arise
from the $A$ equation in the cases \textbf{3.12.2}, \textbf{3.12.3},
\textbf{3.12.4} and \textbf{3.13.5} respectively. Instead equations $_{2}D_{2}$
and $_{3}D_{2}$ arise from the $C$ equation in the cases \textbf{3.13.2}
and \textbf{3.13.3} respectively.
\end{theorem}

To write down the explicit form of the equations in \eqref{eq:h6} 
we used (\ref{rhombiccoeff}, \ref{trapcoeff}) and the fact that the following identities holds:
\begin{equation}
    \begin{aligned}
    f_{n,m} = \Fppp,&\quad |f_{n,m} = \Fmpm,
    \\
    \underline{f}_{n,m} = \Fpmm,&\quad |\underline{f}_{n,m} = \Fmmp.
    \end{aligned}
    \label{eq:idfnm}
\end{equation}
As  mentioned in Appendix \ref{nonaut} if we apply this procedure  to an equation of 
 rhombic type we get a result consistent
with \cite{XP}.

\section{Algebraic Entropy test for (\ref{eq:trapezoidalH4}, \ref{eq:h6})}\label{algent}

 \emph{Algebraic Entropy} is used as a test of integrability for
discrete systems. It  measures the degree of growth  of
the iterates of a rational map \cite{Tremblay2001,Viallet2006}.
The classification of lattice equations based on the algebraic
entropy test
\cite{HietarintaViallet2007} is:
\begin{description}
    \item[Linear growth:] The equation is linearizable.
    \item[Polynomial growth:] The equation is integrable.
    \item[Exponential growth:] The equation is chaotic.
\end{description}

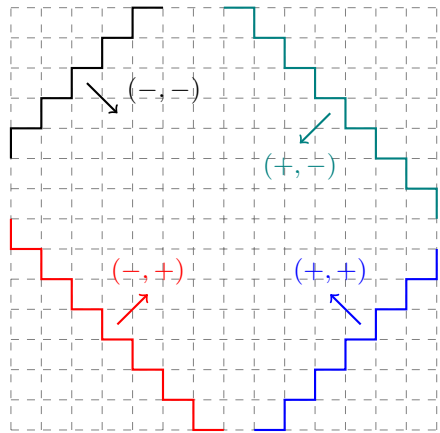
\begin{figure}[hbt]
    \centering
    \begin{tikzpicture}[scale=0.4]
         \draw[style=help lines,dashed] (0,0) grid[step=1cm] (14,14);
         \draw[thick] (0,9)--(0,10)--(1,10)--(1,11)--(2,11)--(2,12)--(3,12)--(3,13)--(4,13)--(4,14)--(5,14);
         \node[above right] at (3+1/2,11-1/2) {$(-,-)$};
         \draw[thick,->] (2+1/2,12-1/2)--(3+1/2,11-1/2); 
         \draw[thick,red] (0,7)--(0,6)--(1,6)--(1,5)--(2,5)--(2,4)--(3,4)--(3,3)--(4,3)--(4,2)--(5,2)--(5,1)--(6,1)--(6,0)--(7,0);
         \node[above,red] at (5-1/2,4+1/2) {$(-,+)$};
         \draw[thick,->,red] (4-1/2,3+1/2)--(5-1/2,4+1/2); 
         \draw[thick,blue] (8,0)--(9,0)--(9,1)--(10,1)--(10,2)--(11,2)--(11,3)--(12,3)--(12,4)--(13,4)--(13,5)--(14,5)--(14,6);
         \node[above,blue] at (11-1/2,4+1/2) {$(+,+)$};
         \draw[thick,->,blue] (12-1/2,3+1/2)--(11-1/2,4+1/2); 
         \draw[thick,teal] (14,7)--(14,8)--(13,8)--(13,9)--(12,9)--(12,10)--(11,10)--(11,11)--(10,11)--(10,12)--(9,12)--(9,13)--(8,13)--(8,14)--(7,14);
         \node[below,teal] at (11-1/2-1,11-1/2-1) {$(+,-)$};
         \draw[thick,->,teal] (11-1/2,11-1/2)--(11-1/2-1,11-1/2-1); 
    \end{tikzpicture}
    \caption{Principal growth directions.}
    \label{fig:princgrowth}
\end{figure}

We have performed the Algebraic Entropy analysis 
in the principal growth directions \cite{Viallet2006} 
as shown in Figure \ref{fig:princgrowth} on all  non-autonomous equations presented in Section \ref{Apollo1} 
to identify their behaviour.
To this end we used the \SymPy~\cite{sympy} module 
\verb!ae2d.py! \cite{GubHay}. We found that a non-autonomous
equation, at difference from what it is assumed in \cite{Viallet2006}, may have
non-constant degrees upon the diagonals. This fact implies that
the equations of Section \ref{Apollo1} do not have in general a single
sequence of degrees.

Let us  explain this result in details. Let us suppose  that we wish to calculate the Algebraic Entropy in
the North-East direction ($-,+$). We should then solve our equation
with respect to $u_{n+1,m+1}$ and then iteratively find the degree of the map in the ($-,+$) direction. According to \cite{Viallet2006} 
one obtains
a matrix, the \emph{evolution matrix},  where in each element we put the degree $d_k$ of the corresponding $k$ lattice point. In  the evolution matrix the degrees
along the diagonals are assumed to be all equal. Then
the evolution matrix will have  the form:
\begin{equation}
    \begin{array}{cccccc}
        1 & d_{2} & d_{3} & d_{4} & d_{5} & d_{6}
        \\
        1 & 1 & d_{2} & d_{3} & d_{4} & d_{5}
        \\
        & 1 & 1 & d_{2} & d_{3} & d_{4}
        \\
        & & 1 & 1 & d_{2} & d_{3}
        \\
        & & & 1 & 1 & d_{2}
        \\
        & & & & 1 & 1
    \end{array}
    \label{vialletdegrees}
\end{equation}
The Algebraic Entropy $\eta$ is defined as:
\begin{equation}
    \eta = \lim_{k\to\infty}\frac{1}{k}\log d_{k}.
    \label{eq:algentdef}
\end{equation}
In our case we obtain experimentally that \emph{the degrees
along the diagonals are not the same}. This means that the actual
evolution matrix will have the form:
\begin{equation}
    \begin{array}{cccccc}
        1 & d_{2}^{(5)} & d_{3}^{(4)} & d_{4}^{(3)} & d_{5}^{(2)} & d_{6}^{(1)}
        \\
        1 & 1 & d_{2}^{(4)} & d_{3}^{(3)} & d_{4}^{(2)} & d_{5}^{(1)}
        \\
        & 1 & 1 & d_{2}^{(3)} & d_{3}^{(2)} & d_{4}^{(1)}
        \\
        & & 1 & 1 & d_{2}^{(2)} & d_{3}^{(1)}
        \\
        & & & 1 & 1 & d_{2}^{(1)}
        \\
        & & & & 1 & 1
    \end{array}
    \label{ourdegrees}
\end{equation}
where by $d_k^{(i)}$ we mean the degree of the iterate map proper
to the $i$-th column of the evolution matrix.
This is obviously a more general case than \eqref{vialletdegrees}.
In the framework of \cite{Viallet2006}
the sequence of degrees will have the form \eqref{vialletdegrees} but in our case we get \eqref{ourdegrees}. We shall call the sequence
\begin{equation}
    1, d_2^{(1)}, d_3^{(1)}, d_4^{(1)}, d_5^{(1)},\ldots
    \label{vialletseq}
\end{equation}
the {\it principal sequence} of growth. A sequence as 
\begin{equation}
    1, d_{2}^{(i)}, d_{3}^{(i)}, d_{4}^{(i)}, d_{5}^{(i)},\ldots
    \label{ourseq}
\end{equation}
 with $i=2$ will be a {\it secondary sequence} of growth, for $i=3$ a {\it third}, and so on. 
 In principle any sequence of the form \eqref{ourseq} will define an 
entropy $\eta^{(i)}$:
\begin{equation}
    \eta^{(i)} = \lim_{k\to\infty}\frac{1}{k}\log d_{k}^{(i)}.
    \label{eq:algentdefi}
\end{equation}
To any sequence of the form \eqref{ourseq}
we can associate a \emph{generating function}, i.e. the function
$g^{(i)}(s)$ such that the coefficients of its Taylor expansion care as near as possible to  
 the $d_{k}^{(i)}$:
\begin{equation}
   g^{(i)}(s) = \sum_{k=0}^{\infty} d_{k}^{(i)}s^{i}.
    \label{eq:genfuncdef}
\end{equation}
Usually we assume these generating functions to be rational and then
the Algebraic Entropy can be calculated  as the modulus of the
smallest pole of the generating function:
\begin{equation}
    \eta^{(i)} = \min\left\{|s| \,\,\middle|\,\, \lim_{\sigma\to s}\left|g^{(i)}(\sigma)\right|=\infty\right\}.
    \label{eq:algentgenfunc}
\end{equation}

We built the program \verb!ae2d.py!  to
analyze the evolution matrices and to search for recurring
sequences of growth. It can extract from a (sufficiently
big) evolution matrix the number of sequences of growth and
analyze them. For further details see \cite{GubHay,gubbiotti_thesis}.

Before considering the Algebraic Entropy of  the equations
presented in Section \ref{Apollo1} we  discuss briefly the Algebraic
Entropy for the rhombic \Hvier~equations, which are well known to be
integrable \cite{XP}. Running \verb!ae2d.py! on these equations one finds 
 that they possess only a principal sequence
with the following isotropic  sequence of degrees:
\begin{equation}
   1, 2, 4, 7, 11, 16, 22, 29, 37,\dots.
    \label{eq:degrhomb}
\end{equation}
To this sequence corresponds the generating function:
\begin{equation}
    g=\frac{s^{2} - s + 1}{\left( 1-s \right)^{3}},
\end{equation}
which gives, through the definition $g=\sum_k d_k s^k$,  the asymptotic fit of the degrees:
\begin{equation} \label{5.9}
    d_{k}=\frac{k(k+1)}{2}+1.
\end{equation}
Since the growth is quadratic $\eta=0$. This result is a confirmation by the Algebraic
Entropy approach
of the integrability of the rhombic \Hvier~equations.
Let us note that the growth sequences \eqref{eq:degrhomb} are the same
in the Rhombic \Hvier~equations also when $\varepsilon=0$, i.e. if
we are in the case of the $H$ equations of the ABS classification.

For the trapezoidal \Hvier~and \Hsechs~equations 
the situation is a bit more complicate. Indeed those  equations
have   in every direction two different sequence of growth,
 the principal  and the secondary one, as the coefficients
of the evolution matrix \eqref{ourdegrees} are 2-periodic. 
The most surprising
feature is however that all the sequences of growth
are linear. This means that all such equations are not
only integrable due to the CAC property,
but also linearizable.

\begin{table}[tb]
    \centering
    \[
    \begin{array}{ccccc}
        \toprule
        \text{Equation} & \multicolumn{4}{c}{\text{Growth direction}}
        \\
        &  -,+  &  +,+  &  +,-  &  -,- 
        \\
        \midrule
        _{t}H_{1}^{\varepsilon} &  L_{1} ,  L_{2}  &  L_{1} ,  L_{2}  &  L_{3} ,  L_{4}  &  L_{3} ,  L_{4} 
        \\
        _{t}H_{2}^{\varepsilon} &  L_{5} ,  L_{6}  &  L_{5} ,  L_{6}  &  L_{7} ,  L_{8}  &  L_{7} ,  L_{8} 
        \\
        _{t}H_{3}^{\varepsilon} &  L_{5} ,  L_{6}  &  L_{5} ,  L_{6}  &  L_{7} ,  L_{8}  &  L_{7} ,  L_{8} 
        \\
        D_{1}  &  L_{0} ,  L_{0}  & L_{0} ,  L_{0}  & L_{0} ,  L_{0}  & L_{0} ,  L_{0} 
        \\
        _{1}D_{2}  &  L_{9} ,  L_{10}  & L_{9} ,  L_{10}  & L_{9} ,  L_{10}  & L_{9} ,  L_{10} 
        \\
        D_{3}  &  L_{11} ,  L_{12}  & L_{11} ,  L_{12}  & L_{11} ,  L_{12}  & L_{11} ,  L_{12} 
        \\
        _{1}D_{4}  &  L_{13} ,  L_{8}  & L_{13} ,  L_{8}  & L_{13} ,  L_{8}  & L_{13} ,  L_{8} 
        \\
        _{2}D_{2}  &  L_{14} ,  L_{15}  &  L_{14} ,  L_{15}  &  L_{15} ,  L_{14}  &  L_{15} ,  L_{14} 
        \\
        _{3}D_{2}  &  L_{16} ,  L_{17}  & L_{16} ,  L_{17}  & L_{17} ,  L_{16}  & L_{17} ,  L_{16} 
        \\
        _{2}D_{4}  &  L_{18} ,  L_{8}  & L_{18} ,  L_{8}  & L_{19} ,  L_{20}  & L_{19} ,  L_{20} 
        \\
        \bottomrule
    \end{array}
\]
    \caption{Sequences of growth for the trapezoidal \Hvier~and \Hsechs~equations. The first one
    is the principal sequence, while the second the secondary. All sequences $L_j$, $j=0, \cdots, 20$ are presented in Table 3.}
    \label{tab:growthh4h6}
\end{table}


Instead of presenting the full evolution matrices \eqref{ourdegrees},
which would be very lengthily and obscure, we present two tables with the relevant properties.
The interested reader will find the full matrices  in \cite{gubbiotti_thesis}.
In Table \ref{tab:growthh4h6} we present a summary
of the sequences of growth of both trapezoidal \Hvier
and \Hsechs~equations. The explicit sequence of the degrees
of growth with generating functions, asymptotic fit of
the degrees of growth and entropy is given in Table 3.

Observing Table \ref{tab:growthh4h6} and Table 3 
we may notice the following facts:
\begin{itemize}
    \item The trapezoidal \Hvier~equations are not isotropic: the
        sequences in the ($-,+$) and ($+,+$) directions are different from
        those in the ($+,-$) and ($-,-$) directions. These results reflect the symmetry
        of the equations.
    \item The \Hsechs~equations, except from $_{2}D_{4}$ which has the same behaviour 
        as the trapezoidal \Hvier, are isotropic. 
        Equations $_{2}D_{2}$ and $_{3}D_{2}$  exchange
        the principal and the secondary sequences from the ($-,+$), ($+,+$)
        directions and the ($+,-$), ($-,-$) directions. 
    \item All  growths, except $L_0$, $L_{3}$, $L_{4}$, $L_{7}$, $L_{8}$,
        $L_{12}$ and $L_{17}$, exhibit a highly oscillatory behaviour.
        They have generating functions of the form:
        \begin{equation}
            g(s) = \frac{P(s)}{(s-1)^{2}(s+1)^{2}}, 
            \label{oscigenfunc}
        \end{equation}
        with the polynomial $P(s)\in \Z[s]$. 
        We may write
        \begin{equation}
            g(s) = P_{0}(s)+\frac{P_{1}(s)}{(s-1)^{2}(s+1)^{2}}, 
            \label{oscigenfunc2}
        \end{equation}
        with $P_{0}(s)\in \Z[s]$ of degree less than $P$ and
        $P_{1}(s)=\alpha s^{3}+\beta s^{2} + \gamma s + \delta$.
        Expanding the second term in \eqref{oscigenfunc2} 
        in partial fractions we obtain:
        \begin{equation}
            \begin{aligned}
            g(s) = P_{0}(s)+
            \frac{1}{4}\left[ 
                {\frac {-\alpha-\gamma+\beta+\delta}{ \left( s+1 \right) ^{2}}}
                +{\frac {\alpha+\gamma+\beta+\delta}{ \left( s-1 \right) ^{2}}}
                \right.
                \\
                \left.
                +{\frac {2 \alpha+\beta-\delta}{s-1}}
                +{\frac {2 \alpha-\beta+\delta}{s+1}}\right].
            \end{aligned}
            \label{oscigenfunc3}
        \end{equation}
       Expanding the term in square parentheses
        in Taylor series we find that:
        \begin{equation}
            g(s) = P_{0}(s) + \sum_{k=0}^{\infty}
            \left[ A_{0} + A_{1} (-1)^{k} + A_{2} k + A_{3} (-1)^{k} k \right] s^{k},
            \label{eq:gexp}
        \end{equation}
        with $A_{i}=A_{i}(\alpha,\beta,\gamma,\delta)$ constants.
      This means the $d_k=A_{0} + A_{1} (-1)^{k} + A_{2} k + A_{3} (-1)^{k} k$ 
      for $k>\deg P_{0}(s)$, and therefore it asymptotically solves a fourth 
      order difference equation. 
      As far as we know, even if some example of behaviour
      containing terms like $(-1)^{k}$ are known \cite{HietarintaViallet2007},
      this is the first time that we observe patterns with
      oscillations given by $k\,(-1)^{k}$.
      We conclude noting that the usage of the algebraic
      entropy as integrability indicator is actually justified by the existence of
      of finite order recurrence relations between the degrees $d_{k}$.
      Indeed the existence of such recurrence relations means that from a local
      property (the sequence of degrees) we may infer a global one 
      (chaoticity/integrability/linearizaribilty) \cite{Viallet2015}.
\end{itemize}

\begin{landscape}
    \begin{longtable}[l]{ >{$\displaystyle}c<{$} >{$\displaystyle}c<{$} >{$\displaystyle}c<{$} %
        >{$\displaystyle}c<{$} >{$\displaystyle}c<{$}}
  \caption{Sequences of growth, generating functions, analytic expression of the 
      degrees and entropy for the trapezoidal \Hvier and \Hsechs equations}
  \\
    \toprule
    \text{Name} & \text{Degrees} & \text{Generating function} & \text{Degree fit} & \text{Entropy}
    \\
    & \Set{d_{k}} & g(s) & d_{k} & \eta
    \\
    \midrule
    \endfirsthead
    \multicolumn{5}{c}{\tablename~\thetable{} -- \emph{Continued from previous page}}
    \\
    \toprule
    \text{Name} & \text{Degrees} & \text{Generating function} & \text{Degree fit} & \text{Entropy}
    \\
    & \Set{d_{k}} & g(s) & d_{k} & \eta
    \\
    \midrule
    \endhead
    \midrule \multicolumn{5}{r}{\emph{Continued on next page}} 
    \\ 
    \bottomrule
    \endfoot
    \bottomrule
    \endlastfoot
    L_{0} & 1, 1, 1, 1, 1, 1, 1, 1, 1, 1, 1, 1, 1\dots & \frac{1}{1-s}
    & 1 & 0
    \\
    L_{1} & 1, 2, 2, 5, 3, 8, 4, 11, 5, 14, 6, 17, 7\dots & \frac{s^{3} + 2 s + 1}{(s-1)^{2}(s+1)^{2}}
    & \frac{\left(-1\right)^{k}}{4} \left(- 2 k + 1\right) + k + \frac{3}{4} & 0
    \\
    L_{2} & 1, 2, 4, 3, 7, 4, 10, 5, 13, 6, 16, 7, 19\dots & \frac{- s^{3} + 2 s^{2} + 2 s + 1}{(s-1)^{2}(s+1)^{2}}
    & \frac{\left(-1\right)^{k}}{4} \left(2 k - 1\right) + k + \frac{5}{4} & 0
    \\
    L_{3} & 1, 2, 2, 3, 3, 4, 4, 5, 5, 6, 6, 7, 7\dots & \frac{- s^{2} + s + 1}{s^{3} - s^{2} - s + 1}
    & - \frac{\left(-1\right)^{k}}{4} + \frac{k}{2} + \frac{5}{4} & 0
    \\
    L_{4} & 1, 2, 4, 5, 7, 8, 10, 11, 13, 14, 16, 17, 19\dots & \frac{s^{2} + s + 1}{s^{3} - s^{2} - s + 1}
    & \frac{\left(-1\right)^{k}}{4} + \frac{3 k}{2} + \frac{3}{4} & 0
    \\
    L_{5} & 1, 2, 4, 6, 11, 10, 19, 14, 27, 18, 35, 22, 43\dots 
    & \frac{s^{6} + 4 s^{4} + 2 s^{3} + 2 s^{2} + 2 s + 1}{(s-1)^{2}(s+1)^{2}}
    & \left(-1\right)^{k} \left(k - \frac{5}{2}\right) + 3 k - \frac{5}{2} & 0
    \\
    L_{6} & 1, 2, 4, 7, 8, 15, 12, 23, 16, 31, 20, 39, 24\dots 
    & \frac{3 s^{5} + s^{4} + 3 s^{3} + 2 s^{2} + 2 s + 1}{(s-1)^{2}(s+1)^{2}}
    & \left(-1\right)^{k} \left(- k + \frac{5}{2}\right) + 3 k - \frac{5}{2} & 0
    \\
    L_{7} & 1, 2, 4, 7, 11, 15, 19, 23, 27, 31, 35, 39, 43\dots & \frac{s^{4} + s^{3} + s^{2} + 1}{(s-1)^{2}}
    & 4k -5 & 0
    \\
    L_{8} & 1, 2, 4, 6, 8, 10, 12, 14, 16, 18, 20, 22, 24\dots & {\frac{s^{2} + 1}{(s-1)^{2}}}
    & 2k & 0
    \\
    L_{9} &
    1, 2, 2, 5, 3, 8, 4, 11, 5, 14, 6, 17, 7,\ldots & 
    \frac{s^{3} + 2 s + 1}{\left(s - 1\right)^{2} \left(s + 1\right)^{2}} & 
    \frac{\left(-1\right)^{k}}{4} \left(- 2 k + 1\right) + k + \frac{3}{4} & 
    0
    \\
    L_{10} &
    1, 2, 3, 5, 5, 8, 7, 11, 9, 14, 11, 17, 13,\ldots & 
    \frac{s^{3} + s^{2} + 2 s + 1}{\left(s - 1\right)^{2} \left(s + 1\right)^{2}} & 
    \frac{\left(-1\right)^{k}}{4} \left(- k + 1\right) + \frac{5 k}{4} + \frac{3}{4} & 
    0
    \\
    L_{11} &
    1, 2, 4, 5, 10, 8, 16, 11, 22, 14, 28, 17, 34,\ldots & 
    \frac{3 s^{4} + s^{3} + 2 s^{2} + 2 s + 1}{\left(s - 1\right)^{2} \left(s + 1\right)^{2}} & 
    \frac{\left(-1\right)^{k}}{4} \left(3 k - 5\right) + \frac{9 k}{4} - \frac{3}{4} & 
    0
    \\
    L_{12} &
    1, 2, 4, 5, 7, 8, 10, 11, 13, 14, 16, 17, 19,\ldots & 
    \frac{s^{2} + s + 1}{\left(s - 1\right)^{2} \left(s + 1\right)} & 
    \frac{\left(-1\right)^{k}}{4} + \frac{3 k}{2} + \frac{3}{4} & 
    0 
    \\  
    L_{13} &
    1, 2, 4, 6, 11, 10, 18, 14, 25, 18, 32, 22, 39, \ldots& 
    \frac{4 s^{4} + 2 s^{3} + 2 s^{2} + 2 s + 1}{\left(s - 1\right)^{2} \left(s + 1\right)^{2}} & 
    \frac{3 \left(-1\right)^{k}}{4} \left(k - 2\right) + \frac{11 k}{4} - \frac{3}{2} & 
    0
    \\
    L_{14} &
    1, 2, 3, 3, 6, 4, 9, 5, 12, 6, 15, 7, 18,\ldots & 
    \frac{s^{4} - s^{3} + s^{2} + 2 s + 1}{\left(s - 1\right)^{2} \left(s + 1\right)^{2}} & 
    \frac{\left(-1\right)^{k}}{4} \left(2 k - 3\right) + k + \frac{3}{4} & 
    0
    \\  
    L_{15} &
    1, 1, 3, 3, 6, 5, 9, 7, 12, 9, 15, 11, 18,\ldots & 
    \frac{s^{4} + s^{3} + s^{2} + s + 1}{\left(s - 1\right)^{2} \left(s + 1\right)^{2}} & 
    \frac{k}{4} \left(\left(-1\right)^{k} + 5\right) & 
    0
    \\
    L_{16} &
    1, 2, 3, 2, 5, 2, 7, 2, 9, 2, 11, 2, 13,\ldots & 
    \frac{- 2 s^{3} + s^{2} + 2 s + 1}{\left(s - 1\right)^{2} \left(s + 1\right)^{2}} & 
    \frac{\left(-1\right)^{k}}{2} \left(k - 1\right) + \frac{k}{2} + \frac{3}{2} & 
    0
    \\  
    L_{17} &
    1, 1, 3, 3, 5, 5, 7, 7, 9, 9, 11, 11, 13, \ldots& 
    \frac{s^{2} + 1}{\left(s - 1\right)^{2} \left(s + 1\right)} & 
    \frac{\left(-1\right)^{k}}{2} + k + \frac{1}{2} & 
    0
    \\
    L_{18} &
    1, 2, 4, 5, 11, 9, 19, 13, 27, 17, 35, 21, 43,\ldots & 
    \frac{s^{6} + s^{5} + 4 s^{4} + s^{3} + 2 s^{2} + 2 s + 1}{\left(s - 1\right)^{2} \left(s + 1\right)^{2}} & 
    \left(-1\right)^{k} \left(k - 2\right) + 3 k - 3 & 
    0 
    \\  
    L_{19} &
    1, 2, 4, 6, 11, 10, 19, 14, 27, 18, 35, 22, 43,\ldots & 
    \frac{s^{6} + 4 s^{4} + 2 s^{3} + 2 s^{2} + 2 s + 1}{\left(s - 1\right)^{2} \left(s + 1\right)^{2}} & 
    \left(-1\right)^{k} \left(k - \frac{5}{2}\right) + 3 k - \frac{5}{2} & 0 
    \\  
    L_{20} &
    1, 1, 3, 2, 6, 3, 9, 4, 12, 5, 15, 6, 18,\ldots & 
    \frac{s^{4} + s^{2} + s + 1}{\left(s - 1\right)^{2} \left(s + 1\right)^{2}} & 
    \frac{\left(-1\right)^{k}}{4} \left(2 k - 1\right) + k + \frac{1}{4} & 0 
\end{longtable}
        \label{tab:growthdetails} 
\end{landscape}
\clearpage

\section{Examples}\label{Apollo2}

In this Section one considers in detail the 
${}_{t}H_{1}^\epsilon$ \eqref{eq:tH1e} and ${}_{1}D_{2}$ \eqref{eq:1D2} equations and shows  the explicit form of the quadruple of matrices
coming from the CAC, the non-autonomous equations which give
the consistency on $\Z^{3}$ and the effective Lax pair. Finally we confirm the predictions of the algebraic entropy analysis 
showing how they can be explicitly linearized.

\subsection{Example 1: ${}_{t}H_{1}^\epsilon$}
\label{sec:th1e}

To construct the Lax Pair for  \eqref{eq:tH1e}
we have to deal with \textbf{Case 3.10.1} in \cite{Boll12b}.
The sextuple of equations we consider is:
\begin{subequations}
    \begin{align}
        A &
        \begin{aligned}[t]
        &= \alpha_{2}\left( x - x_{1} \right)\left( x_{2} -x_{12} \right)
        -\alpha_{1} \left( x-x_{2} \right) \left( x_{1} - x_{12} \right)
        \\
        &+\varepsilon^{2}\alpha_{1}\alpha_{2}\left( \alpha_{1}-\alpha_{2} \right),
        \end{aligned}
        \label{eq:6.1a}
        \\
        B &=\left(x-x_{2}\right)\left(x_{3}-x_{23}\right)-\alpha_{2}\left(1+\epsilon^2x_{3}x_{23}\right)=0,
        \label{eq:6.1b}
        \\
        C &=\left(x-x_{1}\right)\left(x_{3}-x_{13}\right)-\alpha_{1}\left(1+\epsilon^2x_{3}x_{13}\right)=0,
        \label{eq:6.1c}
        \\
        \bar{A} &= \alpha_{2}\left( x_{13} - x_{3} \right)\left( x_{123} -x_{23} \right)
        -\alpha_{1} \left( x_{13}-x_{123} \right) \left( x_{3} - x_{123} \right),
        \label{eq:6.1abar}
        \\
        \bar{B} &=\left(x_{1}-x_{12}\right)\left(x_{13}-x_{123}\right)
        -\alpha_{2}\left(1+\epsilon^2x_{13}x_{123}\right)=0,
        \label{eq:6.1bbar}
        \\
        \bar{C} &=\left(x_{2}-x_{12}\right)\left(x_{23}-x_{123}\right)-\alpha_{1}\left(1+\epsilon^2x_{23}x_{123}\right)=0,
        \label{eq:6.1cbar}
    \end{align}
    \label{6.1}
\end{subequations}
In this sextuple \eqref{eq:tH1e} originates from the
$B$ equation.

We now make the following identifications
\begin{equation}
    \begin{array}{rcccc}
        A\colon & x \to u_{p,n} & x_{1} \to u_{p+1,n} & x_{2} \to u_{p,n+1} & x_{12} \to u_{p+1,n+1}
        \\
        B\colon & x \to u_{n,m} & x_{2} \to u_{n+1,m} & x_{3} \to u_{n,m+1} & x_{23} \to u_{n+1,m+1} 
        \\
        C\colon & x \to u_{p,m} & x_{1} \to u_{p+1,m} & x_{3} \to u_{p,m+1} & x_{13} \to u_{p+1,m+1}
    \end{array}
    \label{eq:ident}
\end{equation}
so that in any  equation we can suppress the dependence
on the appropriate parametric variables. On  $\Z^{3}$ we get 
the following triplet of equations:
\begin{subequations}
\begin{align}
    \tilde{A} &
    \begin{aligned}[t]
        &=\alpha_{2}\left(u_{p,n}-u_{p+1,n}\right)\left(u_{p,n+1}-u_{p+1,n+1}\right)
        \\
        &-\alpha_{1}\left(u_{p,n}-u_{p,n+1}\right)\left(u_{p+1,n}-u_{p+1,n+1}\right)
        \\
        &+\epsilon^2\alpha_{1}\alpha_{2}\left(\alpha_{1}-\alpha_{2}\right)
        {F}_{m}^{\left(+\right)},
    \end{aligned}
    \\
    \tilde{B} &
    \begin{aligned}[t]
        &=\left(u_{n,m}-u_{n+1,m}\right)\left(u_{n,m+1}-u_{n+1,m+1}\right)-\alpha_{2}\\
    &-\alpha_{2}\epsilon^2\left({F}_{m}^{\left(+\right)}u_{n,m+1}u_{n+1,m+1}
    +{F}_{m}^{\left(-\right)}u_{n,m}u_{n+1,m}\right),
    \end{aligned}
    \\
    \tilde{C} &
    \begin{aligned}[t]
    &=\left(u_{p,m}-u_{p+1,m}\right)\left(u_{p,m+1}-u_{p+1,m+1}\right)-\alpha_{1}
    \\
    &-\alpha_{1}\epsilon^2\left({F}_{m}^{\left(+\right)}u_{p,m+1}u_{p+1,m+1}
    +{F}_{m}^{\left(-\right)}u_{p,m}u_{p+1,m}\right).
    \end{aligned}
\end{align}
    \label{eq:tripleth1e}
\end{subequations}
Then with the usual method we find the following Lax Pair:
\begin{subequations}
    \begin{align}
        \tilde L_{n,m}&
        \begin{aligned}[t]
        &=\left(\begin{array}{cc}
        u_{n,m+1} & -u_{n,m}u_{n,m+1}+\alpha_{1}\\
        1 & -u_{n,m}\end{array}\right)
        \\
        &-\epsilon^2\alpha_{1}\left(\begin{array}{cc}
        -{F}_{m}^{\left(-\right)}u_{n,m} & 0\\
        0 & {F}_{m}^{\left(+\right)}u_{n,m+1}\end{array}\right),
        \end{aligned}
        \\
        \tilde M_{n,m}&
        \begin{aligned}[t]
        &=\left(\begin{array}{cc}
        \alpha_{1}\left(u_{n,m}-u_{n+1,m}\right)+\alpha_{2}u_{n+1,m} & -\alpha_{2}u_{n,m}u_{n+1,m}\\
        \alpha_{2} & \alpha_{1}\left(u_{n,m}-u_{n+1,m}\right)-\alpha_{2}u_{n,m}\end{array}\right)
        \\
        &-\epsilon^2\alpha_{1}\alpha_{2}
        \left(\alpha_{1}-\alpha_{2}\right){F}_{m}^{\left(+\right)}\left(\begin{array}{cc}
        0 & 1\\
        0 & 0\end{array}\right).
        \end{aligned}
    \end{align}
    \label{eq:laxpairth1e}
\end{subequations}

Let us now turn to the linearization procedure.
In (\ref{eq:tH1e}) we must set $\alpha_{2}\not=0$, otherwise the equation degenerates into $\left(u_{n,m}-u_{n+1,m}\right)\left(u_{n,m+1}-u_{n+1,m+1}\right)=0$, whose solution is trivial on the lattice. Let us define $u_{n,2k}\doteq w_{n,k}$, $u_{n,2k+1}\doteq z_{n,k}$; then we have the following system of two coupled autonomous difference equations
\begin{subequations}
\bea
\left(w_{n,k}-w_{n+1,k}\right)\left(z_{n,k}-z_{n+1,k}\right)-\epsilon^2\alpha_{2}z_{n,k}z_{n+1,k}-\alpha_{2}=0,\label{Krur1}\\
\left(w_{n,k+1}-w_{n+1,k+1}\right)\left(z_{n,k}-z_{n+1,k}\right)-\epsilon^2\alpha_{2}z_{n,k}z_{n+1,k}-\alpha_{2}=0.\label{Krur2}
\eea
\end{subequations}\\
\noindent Subtracting (\ref{Krur2}) to (\ref{Krur1}), we obtain
\bea \label{20}
\left(w_{n,k}-w_{n+1,k}-w_{n,k+1}+w_{n+1,k+1}\right)\left(z_{n,k}-z_{n+1,k}\right)=0.
\eea\\
\noindent At this point the solution of the system bifurcates:\\
\begin{itemize}
\item {\bf Case 1}: if $z_{n,k}=f_{k}$, where $f_k$ is a generic function of its argument, equation (\ref{20}) is satisfied and  from (\ref{Krur1}) or (\ref{Krur2}) we have that $\epsilon\not=0$ and, solving for $f_{k}$, one gets
\bea \label{21}
f_{k}=\pm\frac{\ri}{\epsilon}.
\eea
\item {\bf Case 2}: if $z_{n,k}\not=f_k$, with $f_k$ given in (\ref{21}),  one has $w_{n,k}=g_{n}+h_{k}$, where $g_n$ and $h_k$ are arbitrary functions of their argument. Hence (\ref{Krur2}) and (\ref{Krur1}) reduce to
\bea
\epsilon^2 z_{n,k}z_{n+1,k}+\kappa_{n}\left(z_{n,k}-z_{n+1,k}\right)+1=0,\ \ \ \kappa_n\doteq\frac{g_{n+1}-g_{n}}{\alpha_{2}},\label{Krur3}
\eea\\
\noindent so that two sub-cases emerge:
\begin{itemize}
\item {\bf Sub-case 2.1}: if $\epsilon=0$, (\ref{Krur3}) then $\kappa_{n}\not=0$, so that, solving,
\bea
z_{n+1,k}-z_{n,k}=\frac{1}{\kappa_{n}},
\eea\\
we get
\bea
z_{n,k}=\left\{\begin{array}{c}
j_{k}+\sum_{l=n_{0}}^{n-1}\frac{1}{\kappa_{l}},\ \ \ n\geq n_{0}+1,\\
j_{k}-\sum_{l=n}^{n_{0}-1}\frac{1}{\kappa_{l}},\ \ \ n\leq n_{0}-1,\end{array}\right.
\eea\\
\noindent where $j_k=z_{n_{0},k}$ is a generic integration function of its argument.
\item {\bf Sub-case 2.2}: if $\epsilon\not=0$, (\ref{Krur3}) is a discrete Riccati equation which can be linearized by the M$\ddot{o}$bius transformation $z_{n,k}\doteq\frac{\ri}{\epsilon}\frac{y_{n,k}-1}{y_{n,k}+1}$ to
\bea \label{25}
\left(\ri \kappa_{n}-\epsilon\right)y_{n+1,k}=\left(\ri \kappa_{n}+\epsilon\right)y_{n,k},
\eea\\
\noindent which, as $\kappa_{n}\not=\pm\ri\epsilon$ because otherwise $y_{n,k}=0$ and $z_{n,k}=-\ri/\epsilon$.  Eq. (\ref{25}) implies
\bea
y_{n,k}=\left\{\begin{array}{c}
j_{k}\prod_{l=n_{0}}^{n-1}\frac{\ri \kappa_{l}+\epsilon}{\ri \kappa_{l}-\epsilon},\ \ \ n\geq n_{0}+1,\\
j_{k}\prod_{l=n}^{n_{0}-1}\frac{\ri \kappa_{l}-\epsilon}{\ri \kappa_{l}+\epsilon},\ \ \ n\leq n_{0}-1,\end{array}\right.
\eea\\
\noindent where $j_k=y_{n_{0},k}$ is another arbitrary integration function of its argument.
\end{itemize}
\end{itemize}
\noindent In conclusion we have always completely integrated the original system.\\

Let us note that in the case $\epsilon=0$ eq. \eqref{eq:tH1e} becomes
\begin{equation}
\left(u_{n,m}-u_{n+1,m}\right)\left(u_{n,m+1}-u_{n+1,m+1}\right)-\alpha_{2}=0,\label{eq:tH1e0}
\end{equation}
so that the contact M\"obius-type transformation
\begin{equation}
u_{n+1,m}-u_{n,m}=\sqrt{\alpha_{2}}\frac{1-w_{n,m}}{%
1+w_{n,m}},\label{eq:mobh1_0lin}
\end{equation}
brings (\ref{eq:tH1e}) into the following first order linear equation:
\begin{equation}
w_{n,m+1}+w_{n,m}=0,\label{eq:h1_0linmob}
\end{equation}
so that:
\begin{equation}
u_{n,m}=\left\{\begin{array}{c}
k_{m}+\sqrt{\alpha_{2}}\sum_{l=n_{0}}^{l=n-1}\frac{1-\left(-1\right)^m w_{l}}{%
1+\left(-1\right)^m w_{l}},\ \ \ n\geq n_{0}+1,\label{eq:th1e0}\\
k_{m}-\sqrt{\alpha_{2}}\sum_{l=n}^{l=n_{0}-1}\frac{1-\left(-1\right)^m w_{l}}{%
1+\left(-1\right)^m w_{l}},\ \ \ n\leq n_{0}-1.\end{array}\right.
\end{equation}
Here $k_{m}=u_{n_{0},m}$ and $w_{n}$, are two arbitrary integration functions.

\subsection{Example 2: ${}_{1}D_{2}$\newline}

Now we  consider the equation $_{1}D_{2}$ \eqref{eq:1D2}.
We consider the sextuple of equations given by \textbf{Case 3.12.2}
in \cite{Boll12b}:
\begin{subequations}
    \begin{align}
        A&=\delta_{2}x+x_{1}+\left(1-\delta_{1}\right)x_{2}+x_{12}\left(x+\delta_{1}x_{2}\right),
        \\
        B &
        \begin{aligned}[t]
        &= \left( x-x_{3} \right)\left( x_{2}-x_{23} \right)
        \\
        &+\lambda\left[ x+x_{3}-\delta_{1} \left( x_{2}+x_{23} \right) \right]
        + \delta_{1}\lambda,
        \end{aligned}
        \\
        C &
        \begin{aligned}[t]
        &= \left( x -x_{3} \right)\left( x_{1}-x_{13} \right)
        -\lambda\left[ \left( \delta_{1}\delta_{2}+\delta_{1}-1 \right)\left( x+x_{3} \right)
        +\delta_{1}\left( x_{1}+x_{13} \right)\right]
        \\
        &+\delta_{1}\left( \delta_{1}\delta_{2}+\delta_{1}-1 \right)\lambda^{2},
        \end{aligned}
        \\
        \bar{A}&=\delta_{2}x_{3}+x_{13}+\left(1-\delta_{1}\right)x_{23}+x_{123}\left(x+\delta_{1}x_{23}\right),
        \\
        \bar{B} &
        \begin{aligned}
        &= \left( x_{1} -x_{13} \right)\left( x_{12}-x_{123} \right)
        \\
        &+\lambda\left[ 2\delta_{2}\left( \delta_{1}-1 \right) + \left( \delta_{1}-1-\delta_{1}\delta_{2}\right)
        -2\delta_{1}x_{12}x_{123}\right],  
        \end{aligned}
        \\
        \bar{C} &= \left( x_{1}-x_{23} \right)\left( x_{12}-x_{123} \right)
        - \lambda\left( 2\delta_{2} +x_{12} + x_{123} \right).
    \end{align}
\end{subequations}
The triplet of consistent dynamical systems on the $3D$-lattice is:
\bea \label{abc}
&&\tilde A\doteq\\
&&\nonumber\doteq\left({F}_{p+n}^{\left(-\right)}-\delta_{1}{F}_{p}^{\left(+\right)}{F}_{n}^{\left(-\right)}+\delta_{2}{F}_{p}^{\left(+\right)}{F}_{n}^{\left(+\right)}\right)u_{p,n}+\left({F}_{p+n}^{\left(+\right)}-\delta_{1}{F}_{p}^{\left(-\right)}{F}_{n}^{\left(-\right)}+\delta_{2}{F}_{p}^{\left(-\right)}{F}_{n}^{\left(+\right)}\right)u_{p+1,n}+\\
&&\nonumber+\left({F}_{p+n}^{\left(+\right)}-\delta_{1}{F}_{p}^{\left(+\right)}{F}_{n}^{\left(+\right)}+\delta_{2}{F}_{p}^{\left(+\right)}{F}_{n}^{\left(-\right)}\right)u_{p,n+1}+\left({F}_{p+n}^{\left(-\right)}-\delta_{1}{F}_{p}^{\left(-\right)}{F}_{n}^{\left(+\right)}+\delta_{2}{F}_{p}^{\left(-\right)}{F}_{n}^{\left(-\right)}\right)u_{p+1,n+1}+\\
&&\nonumber+\delta_{1}\left({F}_{n}^{\left(-\right)}u_{p,n}u_{p+1,n}+{F}_{n}^{\left(+\right)}u_{p,n+1}u_{p+1,n+1}\right)+{F}_{p+n}^{\left(+\right)}u_{p,n}u_{p+1,n+1}+{F}_{p+n}^{\left(-\right)}u_{p+1,n}u_{p,n+1},\\
&&\nonumber\tilde B\doteq\\
&&\nonumber\doteq\lambda\left\{\left[\left(\delta_{1}-1\right){F}_{n}^{\left(+\right)}-\delta_{1}\right]{F}_{p}^{\left(+\right)}+\left(\delta_{1}-1-\delta_{1}\delta_{2}\right){F}_{p}^{\left(-\right)}{F}_{n}^{\left(-\right)}\right\}\left(u_{n,m}+u_{n,m+1}\right)+\\
&&\nonumber+\lambda\left\{\left[\left(\delta_{1}-1\right){F}_{n}^{\left(-\right)}-\delta_{1}\right]{F}_{p}^{\left(+\right)}+\left(\delta_{1}-1-\delta_{1}\delta_{2}\right){F}_{p}^{\left(-\right)}{F}_{n}^{\left(+\right)}\right\}\left(u_{n+1,m}+u_{n+1,m+1}\right)-\\
&&\nonumber-2\delta_{1}\lambda{F}_{p}^{\left(-\right)}\left({F}_{n}^{\left(-\right)}u_{n,m}u_{n,m+1}+{F}_{n}^{\left(+\right)}u_{n+1,m}u_{n+1,m+1}\right)+\\
&&\nonumber+\left(u_{n,m}-u_{n,m+1}\right)\left(u_{n+1,m}-u_{n+1,m+1}\right)+\delta_{1}\lambda^2{F}_{p}^{\left(+\right)}+2\left(\delta_{1}-1\right)\delta_{2}\lambda{F}_{p}^{\left(-\right)},\\
&&\nonumber\tilde C\doteq\\
&&\nonumber\doteq\lambda\left\{\left[\left(1-\delta_{1}\delta_{2}\right){F}_{p}^{\left(+\right)}-\delta_{1}\right]{F}_{n}^{\left(+\right)}+{F}_{p}^{\left(-\right)}{F}_{n}^{\left(-\right)}\right\}\left(u_{p,m}+u_{p,m+1}\right)+2\delta_{2}\lambda{F}_{n}^{\left(-\right)}\\
&&\nonumber+\lambda\left\{\left[\left(1-\delta_{1}\delta_{2}\right){F}_{p}^{\left(-\right)}-\delta_{1}\right]{F}_{n}^{\left(+\right)}+{F}_{p}^{\left(+\right)}{F}_{n}^{\left(-\right)}\right\}\left(u_{p+1,m}+u_{p+1,m+1}\right)+\\
&&\nonumber+\left(u_{p,m}-u_{p,m+1}\right)\left(u_{p+1,m}-u_{p+1,m+1}\right)+\delta_{1}\left(\delta_{1}-1+\delta_{1}\delta_{2}\right)\lambda^2{F}_{n}^{\left(+\right)}.
\eea
We leave out the Lax pair for ${}_{1}D_{2}$ as they are too complicate to write down and not worth while the effort for the reader. If necessary one can always write them down using the standard procedure outlined above.

Let us now turn to the linearization procedure. Notice that there 
is no combination of the parameters $\delta_{1}$ and $\delta_{2}$ 
such that   \eqref{eq:1D2} becomes non-autonomous. So
we are naturally induced to introduce the following four fields:
\begin{equation}
    \begin{array}{cc}
        w_{s,t}=u_{2s,2t}, & y_{s,t}=u_{2s+1,2t},
        \\
        v_{s,t}=u_{2s,2t+1} & z_{s,t} = u_{2s+1,2t+1}. 
    \end{array}
    \label{eq:fourfields}
\end{equation}
which transform \eqref{eq:1D2} into the following system of four 
coupled autonomous difference equations:
\begin{subequations}\label{6.20}
    \begin{align}
        \left(1-\delta_{1}\right)v_{s,t}+\delta_{2}w_{s,t}
        +y_{s,t}+\left(\delta_{1}v_{s,t}+w_{s,t}\right)z_{s,t} &=0,
        \label{Ur51}
        \\
        \left(1-\delta_{1}\right)v_{s+1,t}+\delta_{2}w_{s+1,t}
        +y_{s,t}+\left(\delta_{1}v_{s+1,t}+w_{s+1,t}\right)z_{s,t} &=0,
        \label{Ur52}
        \\
        \left(1-\delta_{1}\right)v_{s,t}+\delta_{2}w_{s,t+1}
        +y_{s,t+1}+\left(\delta_{1}v_{s,t}+w_{s,t+1}\right)z_{s,t} &=0,
        \label{Ur53}
        \\
        \left(1-\delta_{1}\right)v_{s+1,t}+\delta_{2}w_{s+1,t+1}
        +y_{s,t+1}+\left(\delta_{1}v_{s+1,t}+w_{s+1,t+1}\right)z_{s,t} &=0.
        \label{Ur54}
    \end{align}
\end{subequations}
Let us solve \eqref{Ur51} with respect to $y_{s,t}$:
\begin{equation}
    y_{s,t}=-\left(1-\delta_{1}\right)v_{s,t}-\delta_{2}w_{s,t}
    -\left(\delta_{1}v_{s,t}+w_{s,t}\right)z_{s,t}
    \label{Ur55}
\end{equation}
and let us insert $y_{s,t}$  into
\eqref{Ur52} in order to get an equation solvable for $z_{s,t}$. This is possible {\it iff} $\delta_{1}v_{s,t}+w_{s,t}\neq f_{t}$,
with $f_{t}$ a generic function of $t$,
since in this case the coefficient of $z_{s,t}$ is zero. Then the solution of the system (\ref{6.20})
bifurcates.
\begin{description}
    \item[Case 1] Assume that $\delta_{1}v_{s,t}+w_{s,t}\neq f_{t}$, then
        we can solve  with respect to $z_{s,t}$ the expression obtained inserting 
        \eqref{Ur55} into \eqref{Ur52}. We get:
        \begin{equation}
            z_{s,t}=-\frac{\left(1-\delta_{1}\right)\left(v_{s+1,t}-v_{s,t}\right)
            +\delta_{2}\left(w_{s+1,t}-w_{s,t}\right)}{%
                \delta_{1}\left(v_{s+1,t}-v_{s,t}\right)+w_{s+1,t}-w_{s,t}}.
            \label{Ur56}
        \end{equation}
        Now we can substitute \eqref{Ur55} and \eqref{Ur56} together with their 
        difference consequences into \eqref{Ur53} and \eqref{Ur54} and we get two
        equations for $w_{s,t}$ and $v_{s,t}$:
        \begin{subequations}
            \begin{align}
                 \left( \delta_{{1}}\delta_{{2}}+\delta_{{1}}-1 \right)  &
                 \begin{aligned}[t]
                 &\big[w_{{s+1,t+1}}v_{{s,t+1}}w_{{s,t}}+v_{{s+1,t+1}}w_{{s,t+1}}w_{{s+1,t}}
                 \\
                &-v_{{s+1,t+1}}w_{{s,t+1}}w_{{s,t}}+w_{{s,t+1}}v_{{s,t}}w_{{s+1,t+1}}
                \\
                &+v_{{s,t}}w_{{s+1,t+1}}w_{{s+1,t}}-w_{{s,t+1}}v_{{s+1,t}}w_{{s+1,t+1}}
                \\
                &-{w^{2}_{{s,t+1}}}v_{{s,t}}+{w^{2}_{{s,t+1}}}v_{{s+1,t}}
                \\
                &-w_{{s+1,t+1}}v_{{s,t+1}}w_{{s+1,t}}-v_{{s,t}}w_{{s+1,t+1}}w_{{s,t}}
                \\
                &-v_{{s,t}}w_{{s,t+1}}w_{{s+1,t}}+v_{{s,t}}w_{{s,t+1}}w_{{s,t}}
                \\
                &-\delta_{1} ( v_{{s,t}}v_{{s,t+1}}w_{{s+1,t}}+w_{{s,t+1}}v_{{s,t}}v_{{s,t+1}}
                \\
                &-w_{{s+1,t+1}}v_{{s,t+1}}v_{{s,t}}+w_{{s+1,t+1}}v_{{s,t+1}}v_{{s+1,t}}
                \\
                &+v_{{s,t}}v_{{s+1,t+1}}w_{{s,t}}-v_{{s,t}}v_{{s+1,t+1}}w_{{s+1,t}}
                \\
                &-v_{{s,t}}v_{{s,t+1}}w_{{s,t}}-w_{{s,t+1}}v_{{s+1,t}}v_{{s,t+1}})\big]
                \end{aligned}
                \label{eq:Eq1}
                \\
                \left( \delta_{{1}}\delta_{{2}}+\delta_{{1}}-1 \right)& 
                 \begin{aligned}[t]
                     \big[&{w^{2}_{{s+1,t+1}}}v_{{s+1,t}}+w_{{s+1,t+1}}v_{{s,t+1}}w_{{s+1,t}}
                         \\
                         &-w_{{s+1,t+1}}v_{{s,t+1}}w_{{s,t}}-{w^{2}_{{s+1,t+1}}}v_{{s,t}}
                         \\
                         &-w_{{s,t+1}}v_{{s+1,t}}w_{{s+1,t+1}}+w_{{s,t+1}}v_{{s,t}}w_{{s+1,t+1}}
                         \\
                         &-v_{{s+1,t+1}}w_{{s,t+1}}w_{{s+1,t}}+v_{{s+1,t}}w_{{s,t+1}}w_{{s+1,t}}
                         \\
                         &+v_{{s+1,t+1}}w_{{s,t+1}}w_{{s,t}}-v_{{s+1,t}}w_{{s+1,t+1}}w_{{s+1,t}}
                         \\
                         &+v_{{s+1,t}}w_{{s+1,t+1}}w_{{s,t}}-v_{{s+1,t}}w_{{s,t+1}}w_{{s,t}}+
                         \\
                         &\delta_{1}( -v_{{s+1,t}}v_{{s+1,t+1}}w_{{s+1,t}}+v_{{s+1,t}}v_{{s+1,t+1}}w_{{s,t}}
                         \\
                         &-w_{{s+1,t+1}}v_{{s,t}}v_{{s+1,t+1}}-v_{{s+1,t}}v_{{s,t+1}}w_{{s,t}}
                         \\
                         &-v_{{s+1,t+1}}w_{{s,t+1}}v_{{s+1,t}}+w_{{s+1,t+1}}v_{{s+1,t}}v_{{s+1,t+1}}
                         \\
                         &+v_{{s+1,t}}v_{{s,t+1}}w_{{s+1,t}}+v_{{s+1,t+1}}w_{{s,t+1}}v_{{s,t}})\big]
                \end{aligned}
                \label{eq:Eq2}
            \end{align}
            \label{eq:Eq12}
        \end{subequations}
       If 
        \begin{equation}
            \delta_{1}(1+\delta_{2})=1,
            \label{eq:delta12cond}
        \end{equation}
        then \eqref{eq:Eq12} are identically satisfied.
        If $\delta_{1}\left( 1+\delta_{2} \right)\neq1$,
        adding \eqref{eq:Eq1} and \eqref{eq:Eq2}, we obtain:
        \begin{equation}
            \begin{aligned}
            \left(w_{s,t}-w_{s+1,t}-w_{s,t+1}+w_{s+1,t+1}\right) &\cdot
            \\
            \left(v_{s+1,t}-v_{s,t}\right) &
            \left(\delta_{1}+\delta_{1}\delta_{2}-1\right)=0.
            \end{aligned}
            \label{Ur57}
        \end{equation}
        Supposing $\delta_{1}+\delta_{1}\delta_{2}\neq1$  we can annihilate the first or the second factor.
        If set equal to zero the second factor,  we get
        $v_{s,t}=g_{t}$ with $g_{t}$ arbitrary function of $t$
        alone. Substituting this result into\eqref{eq:Eq1} or \eqref{eq:Eq2} , they are        identically satisfied provided  $g_{t}=g_{0}$, with $g_{0}$ constant. Then
        the only non-trivial case is when $\delta_{1}+\delta_{1}\delta_{2}\neq1$,
        and $v_{s,t}\neq g_{t}$.
        In this case we have that $w_{s,t}$ solves the discrete wave equation, i.e. 
         $w_{s,t}= h_{s}+l_{t}$. Substituting $w_{s,t}$
        into \eqref{eq:Eq12} we get a single equation for $v_{s,t}$:
        \begin{equation}
            \begin{aligned}
                \left( h_{s}-h_{s+1} \right)\big[&\left(v_{s+1,t+1}-v_{s+1,t}\right) \left(h_{s}+l_{t+1}\right)
                \\
                &-\left(v_{s,t+1}-v_{s,t}\right)\left(h_{s+1}+l_{t+1}\right)
                \\
                &+\delta_{1}\left( v_{s,t}v_{s+1,t+1}-v_{s+1,t}v_{s,t+1} \right)\big]=0.
            \end{aligned}
            \label{eq:Eq12_2}
        \end{equation}
        which is identically satisfied if $h_{s}=h_{0}$, with $h_{0}$ a constant.
        Therefore we have a non-trivial cases only if
        $h_{s}\neq h_{0}$.
        \begin{description}
            \item[Case 1.1] We have a great simplification if
                in addition to $h_{s}\neq h_0$ we have $\delta_{1}=0$.
                In this case \eqref{eq:Eq12_2}
               is linear:
                \begin{equation} \label{6.27}
                    \begin{aligned}
                    \left(v_{s+1,t+1}-v_{s+1,t}\right) &\left(h_{s}+l_{t+1}\right)
                    \\
                    &-\left(v_{s,t+1}-v_{s,t}\right)\left(h_{s+1}+l_{t+1}\right)=0.
                    \end{aligned}
                \end{equation}
                Eq. (\ref{6.27}) can be easily integrated twice to give:
                \begin{eqnarray}
                    v_{s,t} = \left\{\begin{array}{c}
j_{s}+\sum_{k=t_{0}}^{t-1}\left(h_{s}+l_{k+1}\right)i_{k},\ \ \ t\geq t_{0}+1,\\
j_{s}-\sum_{k=t}^{t_{0}-1}\left(h_{s}+l_{k+1}\right)i_{k},\ \ \ t\leq t_{0}-1,\end{array}\right.
                    \label{eq:vstd1_0sol}
                \end{eqnarray}
                with $i_{t}$ and $j_{s}=v_{s,t_{0}}$ arbitrary integration functions.
            \item[Case 1.2] Now let us suppose again $h_{s}\neq h_{0}$, 
                $\delta_{1}\neq0$ but let us choose $l_{t}=l_{0}$,
                with $l_{0}$ a constant. Performing the translation 
                $\theta_{s,t}= v_{s,t}+\left(h_{s}+l\right)/\delta_{1}$, 
                from \eqref{eq:Eq12_2} we get:
                \begin{equation}
                    \theta_{s,t}\theta_{s+1,t+1}-\theta_{s+1,t}\theta_{s,t+1}=0.
                    \label{eq:linear12}
                \end{equation}
                Eq. (\ref{eq:linear12}) is linearizable via a Cole-Hopf transformation 
                $\Theta_{s,t} = \theta_{s+1,t}/\theta_{s,t}$
               as $v_{s,t}$ cannot be identically zero.
                This linearization
                yields  the general solution $\theta_{s,t}=S_{s}T_{t}$
                with $S_{s}$ and $T_{t}$ arbitrary functions of their argument.
            \item[Case 1.3] Finally if $h_{s}\neq h$, $\delta_{1}\neq0$ and 
                $l_{t}\neq l_{0}$, we perform the transformation 
                \begin{equation}
                    \theta_{s,t}=\frac{1}{\delta_{1}}\left[\left(l_{t}-l_{t+1}\right)v_{s,t}
                    -h_{s}-l_{t+1}\right].
                    \label{eq:trasl}
                \end{equation}
                Then from \eqref{eq:Eq12_2} we get:
                \begin{equation}
                    \theta_{s,t}\left(1+\theta_{s+1,t+1}\right)
                    -\theta_{s+1,t}\left(1+\theta_{s,t+1}\right)=0,
                    \label{eq:linear13}
                \end{equation}
                which, as $v_{s,t}$ cannot be identically zero,
                is easily linearized via the Cole-Hopf transformation
                $\Theta_{s,t}= (1+\theta_{s,t+1})/\theta_{s,t}$ to 
                $\Theta_{s+1,s}-\Theta_{s,t} = 0$ which yields for $\theta_{s,t}$
                the linear equation:
                \begin{equation}
                        \theta_{s,t+1}-p_{t}\theta_{s,t}+1=0,
                    \label{eq:linear13_2}
                \end{equation}
                where $p_{t}$ is an arbitrary integration function.
                Then the general solution is given by:
                \begin{equation}
                    \theta_{s,t}=\left\{\begin{array}{c}
\left(u_{s}-\sum_{l=t_{0}}^{t-1}\prod_{j=t_{0}}^{l}p_{j}^{-1}\right)\prod_{k=t_{0}}^{t-1}p_{k},\ \ \ t\geq t_{0}+1,\\
u_{s}\prod_{k=t}^{t_{0}-1}p_{k}^{-1}+\sum_{l=t}^{t_{0}-1}\prod_{j=t}^{l}p_{j}^{-1},\ \ \ t\leq t_{0}-1,\end{array}\right.\label{eq:case13sol}\end{equation}
                where $u_{s}=\theta_{s,t_{0}}$ is an arbitrary integration function.
        \end{description}
    \item[Case 2] We now suppose:
        \begin{equation}
            w_{s,t} = f_{t} - \delta_{1} v_{s,t},
            \label{eq:case2assmpt}
        \end{equation}
        where $f_{t}$ is a generic function of its argument. 
        Inserting (\ref{Ur55}) and (\ref{eq:case2assmpt}) and their difference 
        consequences into (\ref{6.20}), we get:
        \begin{equation}
            \left(v_{s+1,t}-v_{s,t}\right)\left(\delta_{1}+\delta_{1}\delta_{2}-1\right)=0,
            \label{eq:case2fact}
        \end{equation}
        and the two relations:
        \begin{subequations}
            \begin{align}
                &\begin{aligned}
                f_{t+1} z_{s,t+1} 
                &+ \left[\delta_{1}\left( v_{s,t}-v_{s,t+1} \right)-f_{t+1}\right] z_{s,t}
                \\
                &+\left( \delta_{1}-1 \right) \left( v_{s,t}-v_{s,t+1} \right),
                \end{aligned}
                \label{eq:Eq1tilde}
                \\
                &\begin{aligned}
                f_{t+1} z_{s,t+1} 
                &+ \left[\delta_{1}\left( v_{s+1,t}-v_{s+1,t+1} \right)-f_{t+1}\right] z_{s,t}
                \\
                &+\left( \delta_{1}-1 \right) \left( v_{s+1,t}-v_{s,t+1} \right)
                +\delta_{1}\delta_{2} \left( v_{s+1,t+1}-v_{s,t+1} \right).
                \end{aligned}
                \label{eq:Eq2tilde}
            \end{align}
            \label{eq:Eq12tilde}
        \end{subequations}
        Hence   in \eqref{eq:case2fact} we have a biforcation.
        \begin{description}
            \item[Case 2.1] If we  annihilate the
                second factor in \eqref{eq:case2fact} we get
                $\delta_{1}\left(1+\delta_{2}\right)=1$, i.e.  $\delta_{1}\not=0$. Then adding 
                \eqref{eq:Eq1tilde} and \eqref{eq:Eq2tilde} we obtain:
                \begin{equation}
                    \left(v_{s,t}-v_{s+1,t}-v_{s,t+1}+v_{s+1,t+1}\right)
                    \left(1-\delta_{1}+\delta_{1}z_{s,t}\right)=0.
                    \label{eq:bif21}
                \end{equation}
                It seems that we are facing a new bifurcation.
                However annihilating the second factor, i.e. assuming that 
                $z_{s,t} = 1- 1/\delta_{1}$ give  a trivial case,
                since  \eqref{eq:Eq12tilde} are identically
                satisfied. Therefore we may assume that
                $z_{s,t}\neq1-1/\delta_{1}$. This implies that
                $v_{s,t}=h_{s}+k_{t}$, where $h_{s}$ and $k_{t}$ 
                are generic integration functions of their argument. 
                Inserting it in \eqref{eq:Eq12tilde} we can obtain the following 
                linear equation for $z_{s,t}$:
                \begin{equation}
                    f_{t+1}z_{s,t+1}
                    +\left(\delta_{1}j_{t}-f_{t+1}\right)z_{s,t}
                    +\left(\delta_{1}-1\right)j_{t}=0,
                    \label{Ur59}
                \end{equation}
               with $j_{t} = k_{t+1}-k_{t}$.
                This equation can be solved it gives:
                \begin{equation}
                    \begin{aligned}
                    z_{s,t} &= \left( -1 \right) ^{t} \left( {  \delta_{1}}-1 \right) 
                    \prod _{{t'}=0}^{t-1}{\frac {{\delta_{1}} j_{t'} -f_{t'+1} }{%
                        f_{t'+1}}}\cdot
                    \\
                    &\phantom{=}\sum _{{  t''}=0}^{t-1} 
                    \frac{j_{t''}\left( -1 \right) ^{{t''}}}{%
                    \displaystyle
                    f_{t''+1}\prod _{{t'}=0}^{{  t''}}{\frac {{\delta_{1}} j_{  t'} -
                    f_{t'+1} }{f_{  t'+1} }}}
                    \\
                    &+ \left( -1 \right) ^{t}z_{s,0} \prod_{{t'}=0}^{t-1}
                    {\frac {{  \delta_{1}} j_{  t'} -f_{t'+1}}{f_{  t'+1} }}
                    \end{aligned}
                    \label{eq:case21sol}
                \end{equation}
        \item[Case 2.2] Now we annihilate the first factor in \eqref{eq:case2fact} i.e. 
            $\delta_{1}\left(1+\delta_{2}\right)\neq1$ and $v_{s,t}=l_{t}$, 
            where $l_t$ is an arbitrary function of its argument. From 
            \eqref{eq:Eq12tilde} we obtain  (\ref{Ur59}) with
            $j_{t} = l_{t+1}-l_{t}$.
    \end{description}
\end{description}
In conclusion we have always integrated the original system using
an explicit linearization  through a series of 
transformations and biforcations.

\vspace{12pt}
\noindent 
As a final remark we observe that every transformation used in
the linearization procedure both for the $_{t}H_{1}^{\varepsilon}$
\eqref{eq:tH1e} equation and for the $_{1}D_{2}$ \eqref{eq:1D2}
equation is bi-rational in the fields and their shifts (like
the Cole-Hopf-type transformations). 
This, in fact, has to be expected, since the Algebraic Entropy
test is valid only if we allow transformations which preserve
the algebrogeometric structure underlying the evolution procedure
\cite{Viallet2015}. Indeed there are examples on one-dimensional
lattice of equations chaotic according to the Algebraic Entropy,
but linearizable using some transcendental transformations \cite{GRV}.
So exhibiting the explicit linearization and showing
that it can be attained by bi-rational transformations
is indeed a very strong evidence of the Algebraic Entropy
conjecture \cite{HietarintaViallet2007}. 
Indeed this does not prevent the fact that in some cases such
equations can be linearized through some transcendental transformations.
In fact if $\varepsilon=0$ the $_{1}H_{1}^{\varepsilon}$ equation \eqref{eq:tH1e} can be linearized through the transcendental 
contact transformation:
\begin{equation}
u_{n,m}-u_{n+1,m}=\sqrt{\alpha_{2}} e^{z_{n,m}},\label{31}
\end{equation}
i.e.
\begin{equation}
z_{n,m}=\log\frac{u_{n,m}-u_{n+1,m}}{\sqrt{\alpha_{2}}},\nonumber
\end{equation}
with $\log$ standing for the principal value of the complex logarithm (the principal value is intended for the square root too). The transformation (\ref{31}) brings (\ref{eq:tH1e}) into the following family of first order linear equations:
\bea\label{32}
z_{n,m+1}+z_{n,m}= 2\ri\pi\kappa,\ \ \ \kappa=0,1.
\eea 
However this kind of transformation does not prove
the result of the Algebraic Entropy and the method
explained in Section \ref{sec:th1e} should be considered
the correct one. 

\section{Conclusions} \label{concl}

\indent In this paper we have firstly identified all  independent nonlinear, consistent quad-equations on a single cell not of type $Q$ or rhombic $H^{4}$ up to $(\Mob)^{4}$ transformations on the fields and rotations, translations and inversions of the reference frame.\\
\indent Then we have established an automorphism between the group of $(\Mob)^{4}$ and what we have called the \emph{nonautonomous lifting of} $(\Mob)^{4}$. This result enabled us to construct for each of the quad-equation on a single cell the corresponding dynamical system on a 2D--lattice according to the Boll extension procedure. Then we present all  independent, nonlinear 2D--dynamical systems which are consistent on the 3D--lattice up to the stated lifting,  rotations, translations and inversions of the discrete indexes.\\
\indent For any dynamical system we performed an algebraic entropy analysis which suggested the possibility to linearize all the listed non autonomous equations. Two examples were worked in detail to confirm the algebraic entropy findings.\\
\indent We are working now to show the effective linearizability and Darboux integrability of all the systems we have listed \cite{gls_pavel}. Among the open questions is the use that can be made of the Lax pairs: can it be of some help to show linearization? Moreover, is it possible to demonstrate they are indeed fake Lax pairs\cite{bh1,bh2}? Some results in this direction for the ${}_{t}H_{1}^\epsilon$ equation are presented in \cite{gls16}.

\subsection*{Acknowledgments}
CS and DL  have been partly supported by the Italian Ministry of Education and Research, 2010 PRIN {\it Continuous and discrete nonlinear integrable evolutions: from water waves to symplectic maps}.

GG and DL are supported   by INFN   IS-CSN4 {\it Mathematical Methods of Nonlinear Physics}.

$\\ \\$

\noindent {\bf \Large Appendix} 
\begin{appendices}
\section{Construction of the lattice equation and of its Lax pair}
\label{nonaut}

Let a consider a quad equation $Q=Q(x,x_{1},x_{2},x_{12};\alpha_1,\alpha_2)$ of the ones introduced in {\bf Theorem} 1.
An equation of this kind, even if  consistent around the cube, if not treated with care
can turn into a non integrable equation. 
For example, let us consider  the deformed $H_1$ equation,  $_r H_1^{\varepsilon}$ \cite{ABS2009}:
\begin{equation}
    (x-x_{12})(x_{1}-x_{2}) - (\alpha_{1}-\alpha_{2})(1+\varepsilon^{2} x_{1}x_{2}) = 0,
    \label{h1eps}
\end{equation}
and let us consider the trivial embedding of such equation into a lattice given by the identification\footnote{Generally
from now we will call the field variables $u$ to distinguish them from the ``static'' vertex indices $x$}:
\begin{equation} 
    x \to u_{n,m}, \quad x_{1} \to u_{n+1,m}, \quad x_{2} \to u_{n,m+1},
    \quad x_{12} \to u_{n+1,m+1}.
    \label{trivemb}
\end{equation}
Eq. (\ref{h1eps}) with the identification (\ref{trivemb}) becomes the following lattice equation:
\begin{equation}
    (u_{n,m}-u_{n+1,m+1})(u_{n+1,m}-u_{n,m+1})-(\alpha_{1}-\alpha_{2})(1+\varepsilon^{2}u_{n+1,m}u_{n,m+1})=0.
\label{h1eps_al}
\end{equation}
We apply the algebraic entropy test \footnote{For more details on degree of growth, algebraic entropy and related
    subjects see Section \ref{algent} and references therein.} to
 (\ref{h1eps_al}) and we find the following  growth in the Nord-East ($-,+$) direction of the degrees:  
\begin{equation}
\Set{d_{-,+}}=\Set{1,2,4,9,21,50,120,289\dots}.
\label{h1eps_al_growth}
\end{equation}
This sequence has generating function 
\begin{equation}
g_{-,+}=\frac{1-s-s^{2}}{s^{3}+s^{2}-3s+1}
\label{h1eps_al_gf}
\end{equation}
and therefore has a non-zero algebraic entropy given by:
\begin{equation}
\eta_{-,+}=\log\left( 1+\sqrt{2} \right),
\label{h1eps_al_ent}
\end{equation}
 corresponding to the entropy of a non-integrable lattice equation.

The identification (\ref{trivemb}) is not the only possible embedding. An   embedding of  
(\ref{h1eps}) into a $\Z^{2}$ lattice is obtained by choosing  an elementary cell
of dimension greater than one as the one depicted in Fig. \ref{fig:elcell}.  
In such a case  (\ref{h1eps}) can be extended to a  lattice 
and we will get a different partial difference equation. Since
\emph{a priori}  $Q\neq |Q \neq \underline{Q} \neq |\underline{Q}$  the obtained lattice will be
a four color lattice, see Fig. \ref{fig:elcell}.
Choosing  the origin on the $\Z^{2}$ lattice in the point $x$ we obtain
a lattice equation of the following form:
\begin{equation}  \label{eqn:dysys2}
    \widetilde{Q}\left[ u \right] =
    \begin{cases}
        Q(u_{n,m},u_{n+1,m},u_{n,m+1},u_{n+1,m+1}) & \text{$n=2k,m=2k$, $k\in\Z$},
        \\
        |Q (u_{n,m},u_{n+1,m},u_{n,m+1},u_{n+1,m+1})& \text{$n=2k+1,m=2k$, $k\in\Z$},
        \\
        \underline{Q} (u_{n,m},u_{n+1,m},u_{n,m+1},u_{n+1,m+1})& \text{$n=2k,m=2k+1$, $k\in\Z$},
        \\
        |\underline{Q} (u_{n,m},u_{n+1,m},u_{n,m+1},u_{n+1,m+1})& \text{$n=2k+1,m=2k+1$, $k\in\Z$},
    \end{cases}
  \end{equation}
We could have constructed  $\widetilde{Q}\left[ u \right]$ starting from any other 
point  in Fig. \ref{fig:elcell} as the origin, but such
equations would differ from each other only by a translation, a rotation
or a reflection.  So in the sense of Theorem \ref{Pentalfa0} they will be equivalent
to \eqref{eqn:dysys2}.

In the case of $_r H_{1}^{\varepsilon}$ we have:
\begin{equation}
    _r \widetilde{ H}_{1}^{\varepsilon} =
    \begin{cases}
        \begin{gathered}
            (u_{n,m}-u_{n+1,m+1})(u_{n+1,m}-u_{n,m+1})\\
            -(\alpha_{1}-\alpha_{2})(1+\varepsilon^{2}u_{n+1,m}u_{n,m+1}),
        \end{gathered}
        & \abs{n}+\abs{m} = 2k, \quad k\in \Z,
        \\
        \begin{gathered}
            (u_{n,m}-u_{n+1,m+1})(u_{n+1,m}-u_{n,m+1})\\
            -(\alpha_{1}-\alpha_{2})(1+\varepsilon^{2}u_{n,m}u_{n+1,m+1}),
        \end{gathered}
        & \abs{n} + \abs{m} = 2k+1, \quad k \in \Z,
    \end{cases}
    \label{h1lattice}
\end{equation}
where we have used the symmetry properties of the equation $_r H_{1}^{\varepsilon}$.
This result coincide with that presented in \cite{XP}.

We shall 
now  consider   quad equations which satisfy CAC, since we are concerned about integrability.
So let us consider  six-tuples of quad equations (\ref{system}) assigned to the
faces of a 3D cube as displayed in Fig. \ref{fig:cube}.

First let us notice that, without loss of generality, we can assume that,
if $Q$ is the consistent quad equation we are interested in, then
we may assume that $Q$ is the bottom equation i.e. $Q=A$. 
Indeed if we are interested in an equation on the side of the cube of
Fig. \ref{fig:cube} and these
equations are different from $A$ (once made the appropriate substitutions)
we may just rotate it and re-label
the vertices in an appropriate manner, so that our side equation will become the bottom
equation.
In this way following again \cite{Boll11} and taking into account the result stated above we may build an embedding in $\Z^{3}$, whose points we shall label as triples $(n,m,p)$, of the
consistency cube. To this end we reflect the consistency cube with respect to 
the normal of the back and the right side and then complete again with another
reflection, just in the same way we did for the square. 
Using the same notations as in the planar case we see which are
the proper equations which must be put on the sides of
the ``multicube''. Their form can therefore be described as in 
\eqref{eqn:dysys2}.
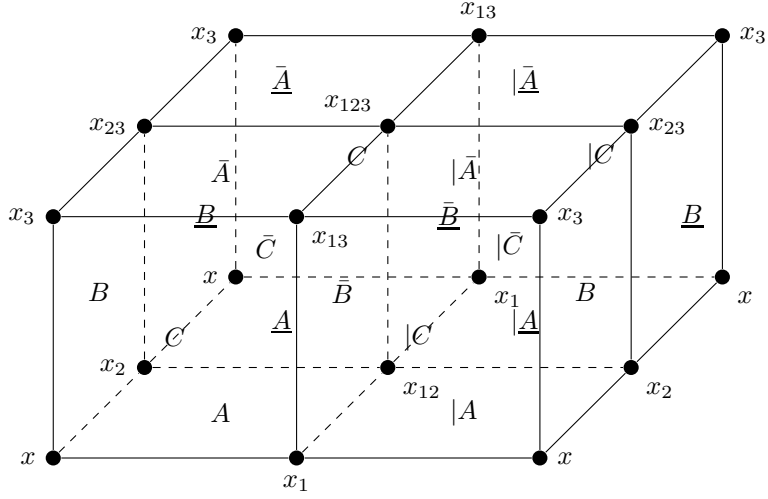
\begin{figure}[htb]
\centering
\begin{tikzpicture}[auto,scale=0.8]
      \node (x) at (0,0) [circle,inner sep=2pt,fill,label=-180:$x$] {};
      \node (x1) at (4,0) [circle,inner sep=2pt,fill,label=-90:$x_{1}$] {};
      \node (x2) at (1.5,1.5) [circle,inner sep=2pt,fill,label=-180:$x_{2}$] {};
      \node (x3) at (0,4) [circle,inner sep=2pt,fill,label=180:$x_{3}$] {};
      \node (x12) at (5.5,1.5) [circle,inner sep=2pt,fill,label=-45:$x_{12}$] {};
      \node (x13) at (4,4) [circle,inner sep=2pt,fill,label=-45:$x_{13}$] {};
      \node (x23) at (1.5,5.5) [circle,inner sep=2pt,fill,label=180:$x_{23}$] {};
      \node (x123) at (5.5,5.5) [circle,inner sep=2pt,fill,label=135:$x_{123}$] {};
      \node (x0r) at (8,0) [circle,inner sep=2pt,fill,label=0:$x$] {};
      \node (x0b) at (3,3) [circle,inner sep=2pt,fill,label=-180:$x$] {};
      \node (x2r) at (9.5,1.5) [circle,inner sep=2pt,fill,label=-45:$x_2$] {};
      \node (x0br) at (11,3) [circle,inner sep=2pt,fill,label=-45:$x$] {};
      \node (x1b) at (7,3) [circle,inner sep=2pt,fill,label=-45:$x_1$] {};
      \node (x3r) at (8,4) [circle,inner sep=2pt,fill,label=0:$x_3$] {};
      \node (x3b) at (3,7) [circle,inner sep=2pt,fill,label=-180:$x_3$] {};
      \node (x23r) at (9.5,5.5) [circle,inner sep=2pt,fill,label=0:$x_{23}$] {};
      \node (x3br) at (11,7) [circle,inner sep=2pt,fill,label=0:$x_3$] {};
      \node (x13b) at (7,7) [circle,inner sep=2pt,fill,label=90:$x_{13}$] {};
      \node (A) at (2.75,0.75) {$A$};
      \node (Ab) at (6.75,0.75) {$|A$};
      \node (bA) at (3.75,2.25) {{$\underline{A}$}};
      \node (bAb) at (7.75,2.25) {{$|\underline{A}$}};
      \node (Aq) at (2.75,4.75) {{$\bar{A}$}};
      \node (Aqb) at (6.75,4.75) {$|\bar{A}$};
      \node (bAq) at (3.75,6.25) {$\bar{\underline{A}}$};
      \node (bAqb) at (7.75,6.25) {$|\bar{\underline{A}}$};
      \node (B) at (0.75,2.75) {$B$};
      \node (Bq) at (4.75,2.75) {$\bar{B}$};
      \node (Br) at (8.75,2.75) {$B$};
      \node (Bb) at (2.50,4.0) {$\underline{B}$};
      \node (Bbq) at (6.50,4.0) {$\bar{\underline{B}}$};
      \node (Bbr) at (10.50,4.0) {$\underline{B}$};
      \node (C) at (2,2) {$C$};
      \node (Cq) at (3.5,3.5) {$\bar{C}$};
      \node (Cb) at (5,5) {$C$};
      \node (Cr) at (6,2) {$|C$};
      \node (Cqr) at (7.5,3.5) {$|\bar{C}$};
      \node (Cbr) at (9,5) {$|C$};
      \draw (x) to (x1);
      \draw (x123) to (x23) to (x3) to (x);
      \draw (x3) to (x13) to (x1);
      \draw (x13) to (x123);
      \draw [dashed] (x) to (x2) to (x12) to (x2r);
      \draw [dashed] (x2) to (x23);
      \draw [dashed] (x2) to (x0b) to (x1b) to (x0br);
      \draw [dashed] (x1) to (x12) to (x1b) to (x13b); 
      \draw [dashed] (x12) to (x123);
      \draw [dashed] (x0b) to (x3b);
      \draw (x13) to (x3r) to (x23r) to (x3br) to (x13b) to (x3b) to (x23);
      \draw (x23r) to (x123) to (x13b);
      \draw (x1) to (x0r) to (x2r) to (x0br) to (x3br);
      \draw (x0r) to (x3r);
      \draw (x2r) to (x23r);
   \end{tikzpicture} 
    \caption{The extension of the consistency cube.}
\label{fig:multicube}
\end{figure}
As a result we end up with Fig. \ref{fig:multicube},
where the functions appearing on the top and on the bottom can be
defined as in \eqref{eqn:dysys2}\footnote{Obviously in the case
of $\bar{A}$ one should  traslate every point by one in the $p$ direction.}
while on the sides we shall have:
\begin{subequations}
\begin{align}
   & \underline{B}(x,x_{2},x_{3},x_{23}) = B(x_{2},x,x_{23},x_{3}),
    \label{Bubar}
    \\
   & \bar{\underline{B}}(x_{1},x_{12},x_{13},x_{123}) =\bar{B}(x_{12},x_{1},x_{123},x_{13}),
        \label{Bbarubar}
        \\
      &  |C(x,x_{1},x_{3},x_{13}) = C(x_{1},x,x_{13},x_{3}),
        \label{Clbar}
        \\
       & |\bar{C}(x_{2},x_{12},x_{23},x_{123}) = \bar{C}(x_{12},x_{2},x_{123},x_{23}).
        \label{Cbarlbar}
    \end{align}
    \label{sides}
\end{subequations}%
From (\ref{sides}) we obtain the analogous in $\Z^{3}$ of 
\eqref{eqn:dysys2}. We 
have a new consistency cube, see Fig. \ref{fig:multicube}, with equations given by\footnote{For the sake of the semplicity of the presentation we have left out if $n$ and $m$ are even or odd integers. They are recovered by comparing with \eqref{eqn:dysys2}.}:
\begin{subequations}
\begin{align}
    \widetilde{\bar{A}}\left[ u \right] &=
    \begin{cases}
        \bar{A}(u_{n,m,p+1},u_{n+1,m,p+1},u_{n,m+1,p+1},u_{n+1,m+1,p+1}),
        \\
        |\bar{A}(u_{n,m,p+1},u_{n+1,m,p+1},u_{n,m+1,p+1},u_{n+1,m+1,p+1}),
        \\
        \bar{\underline{A}}(u_{n,m,p+1},u_{n+1,m,p+1},u_{n,m+1,p+1},u_{n+1,m+1,p+1}),
        \\
        |\bar{\underline{A}}(u_{n,m,p+1},u_{n+1,m,p+1},u_{n,m+1,p+1},u_{n+1,m+1,p+1}),
    \end{cases}
    \label{topA}
    \\
    \widetilde{B}\left[ u \right] &=
    \begin{cases}
        B(u_{n,m,p},u_{n,m+1,p},u_{n,m,p+1},u_{n,m+1,p+1}),
        \\
        \bar{B}(u_{n,m,p},u_{n,m+1,p},u_{n,m,p+1},u_{n,m+1,p+1}),
        \\
        \underline{B}(u_{n,m,p},u_{n,m+1,p},u_{n,m,p+1},u_{n,m+1,p+1}),
        \\
        \bar{\underline{B}}(u_{n,m,p},u_{n,m+1,p},u_{n,m,p+1},u_{n,m+1,p+1}),
    \end{cases}
    \label{sideB}
    \\
    \widetilde{\bar{B}}\left[ u \right] &=
    \begin{cases}
        \bar{B}(u_{n+1,m,p},u_{n+1,m+1,p},u_{n+1,m,p+1},u_{n+1,m+1,p+1}),
        \\
        B(u_{n+1,m,p},u_{n+1,m+1,p},u_{n+1,m,p+1},u_{n+1,m+1,p+1}),
        \\
        \bar{\underline{B}}(u_{n+1,m,p},u_{n+1,m+1,p},u_{n+1,m,p+1},u_{n+1,m+1,p+1}),
        \\
        \underline{B}(u_{n+1,m,p},u_{n+1,m+1,p},u_{n+1,m,p+1},u_{n+1,m+1,p+1}),
    \end{cases}
    \label{sideBbar}
    \\
    \widetilde{C}\left[ u \right] &=
    \begin{cases}
        C(u_{n,m,p},u_{n+1,m,p},u_{n,m,p+1},u_{n+1,m,p+1}),
        \\
        |C(u_{n,m,p},u_{n+1,m,p},u_{n,m,p+1},u_{n+1,m,p+1}),
        \\
        \bar{C}(u_{n,m,p},u_{n+1,m,p},u_{n,m,p+1},u_{n+1,m,p+1}),
        \\
        |\bar{C}(u_{n,m,p},u_{n+1,m,p},u_{n,m,p+1},u_{n+1,m,p+1}),
    \end{cases}
    \label{sideC}
    \\
    \widetilde{\bar{C}}\left[ u \right] &=
    \begin{cases}
        \bar{C}(u_{n,m+1,p},u_{n+1,m+1,p},u_{n,m+1,p+1},u_{n+1,m+1,p+1}),
        \\
        |\bar{C}(u_{n,m+1,p},u_{n+1,m+1,p},u_{n,m+1,p+1},u_{n+1,m+1,p+1}),
        \\
        C(u_{n,m,p},u_{n+1,m+1,p},u_{n,m+1,p+1},u_{n+1,m+1,p+1}),
        \\
        |\bar{C}(u_{n,m+1,p},u_{n+1,m+1,p},u_{n,m+1,p+1},u_{n+1,m,+1p+1}),
    \end{cases}
    \label{sideCbar}
\end{align}
\label{eqn:sideseq}
\end{subequations}
This  means that  the ``multicube'' of Fig.
\ref{fig:multicube} appears as the usual consistency cube of Fig. \ref{fig:cube}
with the following identifications:
\begin{equation}
    A \rightsquigarrow \widetilde{A},
    \quad
    \bar{A} \rightsquigarrow \widetilde{\bar{A}},
    \quad
    B \rightsquigarrow \widetilde{B},
    \quad
    \bar{B} \rightsquigarrow \widetilde{\bar{B}},
    \quad
    C \rightsquigarrow \widetilde{C},
    \quad
    \bar{C} \rightsquigarrow \widetilde{\bar{C}}.
    \label{arrows}
\end{equation}
As an example of this procedure, let us consider again the equation
$_r H_{1}^{\varepsilon}$. $_r H_{1}^{\varepsilon}$ has the  CAC equations:
\begin{subequations}
    \begin{align}
        A &=(x-x_{12})(x_{1}-x_{2}) + (\alpha_{1}-\alpha_{2})(1+\varepsilon x_{1}x_{2}),
        \\
        \bar{A} &=(x_{3}-x_{123})(x_{13}-x_{23}) + (\alpha_{1}-\alpha_{2})(1+\varepsilon x_{3}x_{123}),
        \\
        B &=(x-x_{23})(x_{2}-x_{3}) + (\alpha_{2}-\alpha_{3})(1+\varepsilon x_{2}x_{3}),
        \\
        \bar{B} &=(x_{1}-x_{123})(x_{12}-x_{13}) + (\alpha_{2}-\alpha_{3})(1+\varepsilon x_{1}x_{123}),
        \\
        C &=(x-x_{13})(x_{1}-x_{3}) + (\alpha_{1}-\alpha_{3})(1+\varepsilon x_{1}x_{3}),
        \\
        \bar{C} &=(x_{2}-x_{123})(x_{12}-x_{23}) + (\alpha_{1}-\alpha_{2})(1+\varepsilon x_{2}x_{123}),
    \end{align}
    \label{CACh1eps}
\end{subequations}
therefore from (\ref{eqn:sideseq}, \ref{arrows}) we get the following  consistency on the ``multicube'':
\begin{subequations}
    \begin{align}
    A &=
    \begin{cases}
        \begin{gathered}
            (u_{n,m,p}-u_{n+1,m+1,p})(u_{n+1,m,p}-u_{n,m+1,p})\\
            -(\alpha_{1}-\alpha_{2})(1+\varepsilon^{2}u_{n+1,m,p}u_{n,m+1,p}),
        \end{gathered}
        & \abs{n}+\abs{m} = 2k, \, k\in \Z,
        \\
        \begin{gathered}
            (u_{n,m,p}-u_{n+1,m+1,p})(u_{n+1,m,p}-u_{n,m+1,p})\\
            -(\alpha_{1}-\alpha_{2})(1+\varepsilon^{2}u_{n,m,p}u_{n+1,m+1,p}),
        \end{gathered}
        & \abs{n} + \abs{m} = 2k+1, \, k \in \Z,
    \end{cases}
    \\
    \bar{A} &=
    \begin{cases}
        \begin{gathered}
            (u_{n,m,p+1}-u_{n+1,m+1,p+1})(u_{n+1,m,p+1}-u_{n,m+1,p+1})\\
            -(\alpha_{1}-\alpha_{2})(1+\varepsilon^{2}u_{n,m,p+1}u_{n+1,m+1,p+1}),
        \end{gathered}
        & \abs{n}+\abs{m} = 2k, \, k\in \Z,
        \\
        \begin{gathered}
            (u_{n,m,p+1}-u_{n+1,m+1,p+1})(u_{n+1,m,p+1}-u_{n,m+1,p+1})\\
            -(\alpha_{1}-\alpha_{2})(1+\varepsilon^{2}u_{n+1,m,p+1}u_{n,m+1,p+1}),
        \end{gathered}
        & \abs{n} + \abs{m} = 2k+1, \, k \in \Z,
    \end{cases}
    \\
    B &=
    \begin{cases}
        \begin{gathered}
            (u_{n,m,p}-u_{n,m+1,p+1})(u_{n,m+1,p}-u_{n,m,p+1})\\
            -(\alpha_{2}-\alpha_{3})(1+\varepsilon^{2}u_{n,m+1,p}u_{n,m,p+1}),
        \end{gathered}
        & \abs{n}+\abs{m} = 2k, \, k\in \Z,
        \\
        \begin{gathered}
            (u_{n,m,p}-u_{n,m+1,p+1})(u_{n,m+1,p}-u_{n,m,p+1})\\
            -(\alpha_{2}-\alpha_{3})(1+\varepsilon^{2}u_{n,m,p}u_{n,m+1,p+1}),
        \end{gathered}
        & \abs{n} + \abs{m} = 2k+1, \, k \in \Z,
    \end{cases}
    \\
    \bar{B} &=
    \begin{cases}
        \begin{gathered}
            (u_{n+1,m,p}-u_{n,m+1,p+1})(u_{n+1,m+1,p}-u_{n+1,m,p+1})\\
            -(\alpha_{2}-\alpha_{3})(1+\varepsilon^{2}u_{n+1,m,p}u_{n+1,m+1,p+1},
        \end{gathered}
        & \abs{n}+\abs{m} = 2k, \, k\in \Z,
        \\
        \begin{gathered}
            (u_{n+1,m,p}-u_{n+1,m+1,p+1})(u_{n+1,m+1,p}-u_{n+1,m,p+1})\\
            -(\alpha_{2}-\alpha_{3})(1+\varepsilon^{2}u_{n+1,m+1,p}u_{n+1,m,p+1}),
        \end{gathered}
        & \abs{n} + \abs{m} = 2k+1, \, k \in \Z,
    \end{cases}
    \\
    C &=
    \begin{cases}
        \begin{gathered}
            (u_{n,m,p}-u_{n+1,m,p+1})(u_{n+1,m,p}-u_{n,m,p+1})\\
            -(\alpha_{1}-\alpha_{3})(1+\varepsilon^{2}u_{n+1,m,p}u_{n,m,p+1}),
        \end{gathered}
        & \abs{n}+\abs{m} = 2k, \, k\in \Z,
        \\
        \begin{gathered}
            (u_{n,m,p}-u_{n+1,m,p+1})(u_{n+1,m,p}-u_{n,m,p+1})\\
            -(\alpha_{1}-\alpha_{3})(1+\varepsilon^{2}u_{n,m,p}u_{n+1,m,p+1}),
        \end{gathered}
        & \abs{n} + \abs{m} = 2k+1, \, k \in \Z,
    \end{cases}
    \\
    \bar{C} &=
    \begin{cases}
        \begin{gathered}
            (u_{n,m+1,p}-u_{n+1,m+1,p+1})(u_{n+1,m+1,p}-u_{n,m+1,p+1})\\
            -(\alpha_{1}-\alpha_{3})(1+\varepsilon^{2}u_{n,m+1,p}u_{n+1,m+1,p+1}),
        \end{gathered}
        & \abs{n}+\abs{m} = 2k, \, k\in \Z,
        \\
        \begin{gathered}
            (u_{n,m+1,p}-u_{n+1,m+1,p+1})(u_{n+1,m+1,p}-u_{n,m+1,p+1})\\
            -(\alpha_{1}-\alpha_{2})(1+\varepsilon^{2}u_{n+1,m+1,p}u_{n,m+1,p+1}),
        \end{gathered}
        & \abs{n} + \abs{m} = 2k+1, \, k \in \Z,
    \end{cases}
    \end{align}
    \label{CACh1epslattice}
\end{subequations}

Up to now we showed how, given a CAC quad equation
$Q$,  it is possible to embed it into a partial difference equation in $\Z^{2}$
given by  \eqref{eqn:dysys2}. Furthermore we showed that this procedure
can be extended along the third  dimension in such a way that the consistency
is preserved. This have been done  following \cite{Boll11} and
filling the details (which are going to be important).

The partial difference equation \eqref{eqn:dysys2} is not very manageable
since we have to change equation according to the point of the
lattice we are in. It will be more efficient to have an expression which ``knows''
by itself in which point we are.
This can obtained by going over to non-autonomous equations
as was done in the BW lattice case \cite{XP}.

We shall present here briefly how from \eqref{eqn:dysys2} it is possible to construct an equivalent non-autonomous system, and moreover how to construct the
non-autonomous version of CAC \eqref{eqn:sideseq}.

We take an equation $\hat Q$ constructed by a linear combination
of the equations \eqref{eqn:dysys1} with $n$ and $m$ depending coefficients:
\begin{equation}
    \begin{aligned}
    \widehat{Q} &= f_{n,m}\,Q(u_{n,m},u_{n+1,m},u_{n,m+1},u_{n+1,m+1})
    + \\ &+|f_{n,m}\,|Q (u_{n,m},u_{n+1,m},u_{n,m+1},u_{n+1,m+1})+\\
    &+\underline{f}_{n,m}\,\underline{Q} (u_{n,m},u_{n+1,m},u_{n,m+1},u_{n+1,m+1})
    +\\ &+|\underline{f}_{n,m}\,|\underline{Q} (u_{n,m},u_{n+1,m},u_{n,m+1},u_{n+1,m+1}).
    \end{aligned}
    \label{eqn:dysys3}
\end{equation}
We require  that it satisfies the following conditions:
\begin{enumerate}
    \item The coefficients are periodic of period $2$ in both directions,
        since, in the $\Z^{2}$ embedding,  the elementary cell is
         a $2\times 2$ one.
        \label{cond1}
    \item The coefficients are such that they produce the right equation
        in a given lattice point as specified in \eqref{eqn:dysys2}.\label{cond2}
\end{enumerate}

Condition \ref{cond1} implies that any  function $\tilde f_{n,m}$
in \eqref{eqn:dysys3}, i.e. either $f_{n,m}$ or  $|f_{n,m}$ or $\underline{f}_{n,m}$ or $|\underline{f}_{n,m}$, solves the two ordinary difference equations:
\begin{equation}
    \tilde {f}_{n+2,m}-\tilde{f}_{n,m}=0,\quad \tilde{f}_{n,m+2}-\tilde{f}_{n,m}=0,
    \label{feqs}
\end{equation}
whose solution is:
\begin{equation}
    \tilde {f}_{n,m}= c_{0}+c_{1}\left( -1 \right)^{n}+c_{2}\left( -1 \right)^{m}+c_{3}\left( -1 \right)^{n+m}
    \label{fsol}
\end{equation}
with $c_{i}$ constants to be determined. 

The condition \ref{cond2}  depends on the choice of
the equation in \eqref{eqn:dysys2} and will give some ``boundary
conditions'' for the function $\tilde{f}$, allowing us to fix the coefficients 
$c_{i}$. For $f_{n,m}$, for example,  we have,
substituting the appropriate lattice points, the following conditions:
\begin{equation}
    f_{2k,2k}=1,\quad f_{2k+1,2k}=f_{2k,2k+1}=f_{2k+1,2k+1}=0,
    \label{fcond}
\end{equation}
which yield:
\begin{subequations}
\begin{equation}
    f_{n,m}= \frac{1+\left( -1 \right)^{n}+\left( -1 \right)^{m}+\left( -1 \right)^{n+m}}{4}.
    \label{fsol2}
\end{equation}
In an analogous manner we obtain the form of the other functions
in \eqref{eqn:dysys3}:
\begin{align}
    |f_{n,m} &= \frac{1-\left( -1 \right)^{n}+\left( -1 \right)^{m}-\left( -1 \right)^{n+m}}{4},
    \\
    \underline{f}_{n,m} &= \frac{1+\left( -1 \right)^{n}-\left( -1 \right)^{m}-\left( -1 \right)^{n+m}}{4},
    \\
    |\underline{f}_{n,m} &= \frac{1-\left( -1 \right)^{n}-\left( -1 \right)^{m}+\left( -1 \right)^{n+m}}{4}.
    \label{fsol3}
\end{align}
\label{fsoltot}
\end{subequations}
Then inserting (\ref{fsoltot}) in  \eqref{eqn:dysys3} we obtain  a non-autonomous equation which corresponds to
 \eqref{eqn:dysys2}.

 If the quad-equation $Q$ possess some discrete symmetries, the expression \eqref{eqn:dysys3} 
greatly simplify.
If an equation $Q$ is invariant under the discrete group $D_{4}$ we trivially have, using \eqref{eq:symmsquare},
$ \widehat{Q}=Q$. This result states that an equation with the symmetry (\ref{eq:symmsquare})
is  defined on a monochromatic lattice, as expected since we are in the case of the ABS
classification \cite{ABS03}.
If the equation $Q$ has the symmetries of the \emph{rhombus}, namely (\ref{rhombic}),
we get:
\begin{equation}
    \widehat{Q}=(f_{n,m}+|\underline{f}_{n,m})Q 
    + (|f_{n,m}+\underline{f}_{n,m})|Q
\end{equation}
and using \eqref{fsoltot}:
\begin{equation}
    f_{n,m}+|\underline{f}_{n,m} = F^{(+)}_{n+m},
    \quad
    |f_{n,m}+\underline{f}_{n,m} = F^{(-)}_{n+m}.
    \label{rhombiccoeff}
\end{equation}
where:
\begin{equation}
    F_{k}^{(\pm)} = \frac{1\pm(-1)^{k}}{2}, \quad k \in \Z.
    \label{Gfunc}
\end{equation}
This obviosly match with the results in \cite{XP}.

In the case of \emph{trapezoidal}
symmetry (\ref{trapezoidal1}) one obtains:
\begin{equation}
    \widehat{Q} =(f_{n,m}+|f_{n,m})Q 
    + (\underline{f}_{n,m}+|\underline{f}_{n,m})\underline{Q},
    \label{traplat}
\end{equation}
with
\begin{equation}
        f_{n,m}+|f_{n,m} = F_{m}^{(+)}, \quad
        \underline{f}_{n,m}+|\underline{f}_{n,m} = F_{m}^{(-)},
    \label{trapcoeff}
\end{equation}

As an example of such construction let us consider again
$_r H_{1}^{\varepsilon}$ \eqref{h1eps}.
Since we are in the rhombic case \cite{XP} we  use formula
\eqref{rhombiccoeff} and get:
\begin{equation}
    \begin{aligned}
        _r \widehat{H}_{1}^{\varepsilon} &= 
        (u_{n,m}-u_{n+1,m+1})(u_{n+1,m}-u_{n,m+1})-(\alpha_{1}-\alpha_{2})
        \\
        &+(\alpha_{1}-\alpha_{2})\varepsilon^{2}\left(F^{(+)}_{n+m}\; u_{n+1,m}u_{n,m+1}
        +|F^{(-)}_{n+m} \; u_{n,m}u_{n+1,m+1}\right) = 0,
    \end{aligned}
    \label{h1epshat}
\end{equation}
which corresponds  to the case $\sigma=1$ of \cite{XP} (the discussion of the meaning of parameter $\sigma$ in  \cite{XP} is postponed to the end this Appendix).

The consistency of a generic
system of quad equations is obtained by considering the
consistency of the tilded equations as displayed in 
\eqref{arrows}. We now construct, starting from (\ref{CACh1epslattice}),  the non autonomous partial difference equations in the $(n,m)$
variables using the
 weights $\tilde{f}_{n,m}$, as given in  \eqref{eqn:dysys3}, applied to the relevant equations.   
Carrying out such construction, we  end with the following sextuple of equations:
\begin{subequations}
    \begin{align}
   &&\widehat{A}(u_{n,m,p},u_{n+1,m,p},u_{n,m+1,p},u_{n+1,m+1,p})=  f_{n,m}A +|f_{n,m}|A +
    \\ 
    \nonumber &&\qquad \qquad  +\underline{f}_{n,m}\underline{A}+|\underline{f}_{n,m}|\underline{A} = 0,
    \\
    &&\widehat{\bar{A}} (u_{n,m,p+1},u_{n+1,m,p+1},u_{n,m+1,p+1},u_{n+1,m+1,p+1})=  f_{n,m}\bar{A} +|f_{n,m}|\bar{A} +
    \\ 
    \nonumber &&\qquad \qquad  \qquad \qquad \qquad+\underline{f}_{n,m}\underline{\bar{A}}+|\underline{f}_{n,m}|\underline{\bar{A}} = 0,
    \\
    &&\widehat{B}(u_{n,m,p},u_{n,m+1,p},u_{n,m,p+1},u_{n,m+1,p+1}) =  f_{n,m}B +|f_{n,m}|\bar{B} +
    \\ \nonumber &&\qquad  \qquad \qquad\qquad \qquad +\underline{f}_{n,m}\underline{B}+|\underline{f}_{n,m}|\underline{\bar{B}} =0 ,
    \\
    &&\widehat{\bar{B}} (u_{n+1,m,p},u_{n+1,m+1,p},u_{n+1,m,p+1},u_{n+1,m+1,p+1})=  f_{n,m}\bar{B} +|f_{n,m}|B +
    \\ \nonumber&&\qquad  \qquad \qquad \qquad \qquad+\underline{f}_{n,m}\underline{\bar{B}}+|\underline{f}_{n,m}|\underline{B} =0 ,
    \\
    &&\widehat{C}(u_{n,m,p},u_{n+1,m,p},u_{n,m,p+1},u_{n+1,m,p+1}) =  f_{n,m}C +|f_{n,m}|C +
    \\ \nonumber &&\qquad  \qquad \qquad \qquad \qquad +\underline{f}_{n,m}\bar{C}+|\underline{f}_{n,m}|\bar{C} =0,
    \\
    &&\widehat{\bar{C}} (u_{n,m+1,p},u_{n+1,m,p},u_{n,m+1,p+1},u_{n+1,m+1,p+1})=  f_{n,m}\bar{C} +|f_{n,m}|\bar{C} +
    \\ \nonumber &&\qquad  \qquad \qquad \qquad \qquad+\underline{f}_{n,m}C+|\underline{f}_{n,m}|C = 0,
    \end{align}
    \label{nonautsixtuple}
\end{subequations}
where all the functions on the right hand side of  the equality sign  are  evaluated on the point
indicated on the left hand side.

We note that a Lax pair obtained by making use of equations
\eqref{nonautsixtuple} will be effectively a pair,
since the couples $(\widehat{B},\widehat{\bar{B}})$ and
$(\widehat{C}, \widehat{\bar{C}})$ are related by translation so they are just two different solutions of the same equation.
Indeed by using the properties of the functions $\tilde{f}_{n,m}$
we have:
\begin{equation}
    \widehat{\bar{B}} = T_{n} \widehat{B}, \quad \widehat{\bar{C}} = T_{m}\widehat{C},
    \label{BCtrasl}
\end{equation}
where $T_{n}$ is the operator of translation in the $n$ direction,
and $T_{m}$ the operator of translation in the $m$ direction.
This allow us to construct B\"acklund transformations and  Lax pair 
in the usual way\cite{NW,Bobenko2008book}.

As a final example we shall derive the non-autonomous
side equations for $H_{1}^{\epsilon}$ and its Lax pair.
We will then confront the  result  with that obtained in \cite{XP}.
Considering (\ref{CACh1eps}, \ref{nonautsixtuple}, \ref{BCtrasl}) and using
the fact that the equation is rhombic \eqref{rhombiccoeff}
we get the following result:
\begin{subequations}
    \begin{align}
        \widehat{A} &= 
        \begin{aligned}[t]
            &(u_{n,m,p}-u_{n+1,m+1,p})(u_{n+1,m,p}-u_{n,m+1,p})
            -(\alpha_{1}-\alpha_{2}) \cdot
            \\
            &\cdot\left[1+\varepsilon^{2}\left(F^{(+)}_{n+m}\;u_{n+1,m,p}u_{n,m+1,p}
            +F^{(-)}_{n+m} \;u_{n,m,p}u_{n+1,m+1,p}\right)\right] = 0,
        \end{aligned}
        \\
        \widehat{\bar{A}} &=
        \begin{aligned}[t]
            &(u_{n,m,p+1}-u_{n+1,m+1,p+1})(u_{n+1,m,p+1}-u_{n,m+1,p+1})
            -(\alpha_{1}-\alpha_{2})\cdot
            \\
            &\cdot\left[1+\varepsilon^{2}\left(F^{(-)}_{n+m}\;u_{n+1,m,p+1}u_{n,m+1,p+1}
            +F^{(+)}_{n+m}\; u_{n,m,p+1}u_{n+1,m+1,p+1}\right)\right] = 0,
        \end{aligned}
        \\
        \widehat{B} &= 
        \begin{aligned}[t]
            &(u_{n,m,p}-u_{n,m+1,p+1})(u_{n,m+1,p}-u_{n,m,p+1})
            -(\alpha_{2}-\alpha_{3})\cdot
            \\
            &\cdot\left[1+\varepsilon^{2}\left(F^{(+)}_{n+m}\;u_{n,m+1,p}u_{n,m,p+1}
            +F^{(-)}_{n+m}\; u_{n,m,p}u_{n,m+1,p+1}\right)\right] = 0,
        \end{aligned}
        \\
        \widehat{C} &= 
        \begin{aligned}[t]
            &(u_{n,m,p}-u_{n+1,m,p+1})(u_{n+1,m,p}-u_{n,m,p+1})
            -(\alpha_{1}-\alpha_{3})\cdot
            \\
            &\cdot\left[1+\varepsilon^{2}\left(F^{(+)}_{n+m}\;u_{n+1,m,p}u_{n,m,p+1}
            +F^{(-)}_{n+m} \;u_{n,m,p}u_{n+1,m,p+1}\right)\right] = 0,
        \end{aligned}
    \end{align}
    \label{h1epsnonautCAC}
\end{subequations}
From the equations $\widehat{B}$ and $\widehat{C}$ we find,
up to a sign and commond factor, which we are can
eliminate since the Lax pair is defined from CAC only
up to projective equivalence \cite{HV,BrH},
the following Lax pair:
\begin{subequations}
    \begin{align}
        L&
        \begin{aligned}[t]
            &=
        \begin{pmatrix}
            u_{n,m}
            & 
            \alpha_{1} -  \alpha_{3} -  u_{n,m} u_{n+1,m}
            \\
            1 & -u_{n+1,m}
        \end{pmatrix}
        \\
        &+(\alpha_{1}-\alpha_{3})\varepsilon^{2}
        \begin{pmatrix}
            \Fp{n+m} u_{n+1,m} & 0
            \\
            0 & -\Fm{n+m} u_{n,m}
        \end{pmatrix}
        \end{aligned}
        \label{}
        \\
        M &
        \begin{aligned}[t]
        &=\begin{pmatrix}
            u_{n,m} & \alpha_{2} - \alpha_{3} -u_{n,m} u_{n,m+1}
            \\
            1 & -u_{n,m+1}
        \end{pmatrix}
        \\
        &+\left( \alpha_{2}-\alpha_{3} \right)\varepsilon^{2}
        \begin{pmatrix}
            \Fp{n+m} u_{n,m+1} & 0
            \\
            0 & -\Fm{n+m} u_{n,m}
        \end{pmatrix},
        \end{aligned}
        \\
        \bar{L}&
        \begin{aligned}[t]
            &=
        \begin{pmatrix}
            u_{n,m+1}
            & 
            \alpha_{1} -  \alpha_{3} -  u_{n,m+1} u_{n+1,m+1}
            \\
            1 & -u_{n+1,m+1}
        \end{pmatrix}
        \\
        &+(\alpha_{1}-\alpha_{3})\varepsilon^{2}
        \begin{pmatrix}
            \Fm{n+m} u_{n+1,m+1} & 0
            \\
            0 & -\Fp{n+m} u_{n,m+1}
        \end{pmatrix}
        \end{aligned}
        \\
        \bar{M}&
        \begin{aligned}[t]
        &=\begin{pmatrix}
            u_{n+1,m} & \alpha_{2} - \alpha_{3} -u_{n+1,m} u_{n+1,m+1}
            \\
            1 & -u_{n+1,m+1}
        \end{pmatrix}
        \\
        &+\left( \alpha_{2}-\alpha_{3} \right)\varepsilon^{2}
        \begin{pmatrix}
            \Fm{n+m} u_{n+1,m+1} & 0
            \\
            0 & -\Fp{n+m} u_{n+1,m}
        \end{pmatrix}.
        \end{aligned}
    \end{align}
    \label{lph1eps}
\end{subequations}
This is a Lax pair since $\bar{L}=T_{m}L$ and $\bar{M} = T_{n} M$.
We find that such matrices give as compatibility the equation \eqref{h1eps}
and correspond to the following proportionality factor $\tau$  \cite{BrH}:
\begin{equation}
    \tau = \frac{1 +\varepsilon^{2} \left( F^{(-)}_{n+m} u_{n,m}+ F^{(+)}_{n+m} u_{n+1,m} \right)}{%
        1+\varepsilon^{2}\left( F^{(-)}_{n+m} u_{n,m} + F^{(+)}_{n+m}u_{n,m+1} \right)}.
    \label{tauh1eps}
\end{equation}
The Lax pair \eqref{lph1eps} is Gauge equivalent to that obtained in \cite{XP}
with  gauge:
\begin{equation}
    G = 
    \begin{pmatrix}
        0 & 1
        \\
        -1 & 0
    \end{pmatrix}.
    \label{gaugeh1eps}
\end{equation}

A similar calculation could be done for the other two rhombic
equations, and, up to gauge transformations, will give the same result. Indeed
the gauge transformations \eqref{gaugeh1eps} is needed
for $H_{1}^{\varepsilon}$ and $H_{2}^{\varepsilon}$ whereas
for $H_{3}^{\varepsilon}$ we need the gauge:
\begin{equation}
    \tilde{G} =
    \begin{pmatrix}
        0 & (-1)^{n+m}
        \\
        -(-1)^{n+m} & 0
    \end{pmatrix}.
    \label{gaugeh3eps}
\end{equation}

We end this Appendix with the following remark: at the level of
the non-autonomous equations the choice of origin of $\Z^{2}$ in a
point different from $x$ in \eqref{eqn:dysys2} would have led to different 
initial conditions in \eqref{fcond}, which ultimately lead to the following
form for the functions $f$:
\begin{subequations}
    \begin{align}
    f_{n,m}^{\sigma_{1},\sigma_{2}} 
    &= \frac{1+\sigma_{1}\left( -1 \right)^{n}+\sigma_{2}\left( -1 \right)^{m}
    +\sigma_{1}\sigma_{2}\left( -1 \right)^{n+m}}{4},
    \\
    |f_{n,m}^{\sigma_{1},\sigma_{2}} 
    &= \frac{1-\sigma_{1}\left( -1 \right)^{n}+\sigma_{2}\left( -1 \right)^{m}
    -\sigma_{1}\sigma_{2}\left( -1 \right)^{n+m}}{4},
    \\
    \underline{f}_{n,m}^{\sigma_{1},\sigma_{2}} 
    &= \frac{1+\sigma_{1}\left( -1 \right)^{n}-\sigma_{2}\left( -1 \right)^{m}
    -\sigma_{1}\sigma_{2}\left( -1 \right)^{n+m}}{4},
    \\
    |\underline{f}_{n,m}^{\sigma_{1},\sigma_{2}} 
    &= \frac{1-\sigma_{1}\left( -1 \right)^{n}-\sigma_{2}\left( -1 \right)^{m}
    +\sigma_{1}\sigma_{2}\left( -1 \right)^{n+m}}{4},
    \end{align}%
    \label{eqn:fsolsigma12}%
\end{subequations}
where the two constants $\sigma_{i}\in\Set{\pm1}$ depends on the point
chosen. Indeed if the point is $x$ we have $\sigma_{1}=\sigma_{2}=1$,
whereas if we choose $x_{1}$ we have $\sigma_{1}=1$, $\sigma_{2}=-1$,
if we choose $x_{2}$ then $\sigma_{1}=-1$ and $\sigma_{2}=1$ and finally
if we choose $x_{12}$ we shall put $\sigma_{1}=\sigma_{2}=-1$.
It is easy to see that in the rhombic and in the trapezoidal 
case the functions \eqref{eqn:fsolsigma12} collapse to the $\sigma$ version of the 
functions $F_{k}^{(\pm)}$ as given by \eqref{Gfunc}:
\begin{equation}
   F_{k}^{(\pm,\sigma)} = \frac{1\pm\sigma(-1)^{k}}{2}.
    \label{Ffuncsigma}
\end{equation}
The final equation will then depend on $\sigma_{1}$ or $\sigma_{2}$,
 only if rhombic or trapezoidal.

It was proved in \cite{XP} that the transformations:
\begin{equation}
    v_{n,m} = u_{n+1,m}, \quad w_{n,m} = u_{n,m+1},
    \label{eq:backxp}
\end{equation}
map a rhombic equation with a certain $\sigma$ into
the same rhombic equation with $-\sigma$. In \cite{XP}
this fact was used to construct a Lax Pair and B\"acklund
transformations. An analogous result can be easily proven
for trapezoidal equations: using the transformation $w_{n,m} = u_{n,m+1}$
we can send a trapezoidal equation with a certain $\sigma$ into
the same equation with $-\sigma$. A similar trasformation in the
$n$ direction would just trivally leave invariant the trapezoidal
equation, since there is no explicit dependence on $n$.
However in general, if an equation does not posses discrete symmetries, 
as it is the case for a \Hsechs~equation, no trasformation like \eqref{eq:backxp}
would take the equation into itself with different coefficents.
We can anyway construct  a Lax pair with the
procedure explained above, which is then slightly more general
than the approach based on the transformations \eqref{eq:backxp}.
\end{appendices}


\end{document}